\documentclass{aa}  

\usepackage{graphicx}
\usepackage{txfonts}
\usepackage{color}
\usepackage{lscape}
\begin{document}

   \title{Evolved stars and the origin of abundance trends in planet hosts
 \thanks{
   Based on observations made with the Mercator Telescope; 
   on observations made with  the Nordic Optical Telescope; 
   on observations made with the Italian Telescopio Nazionale Galileo; %
   on observations collected  at the Centro Astron\'omico Hispano
   Alem\'an (CAHA) at Calar Alto; 
   and on data products from observations made with ESO Telescopes at the La Silla Paranal Observatory
   under programme ID 
   072.C-0488(E),
   080.D-0347(A),
   081.D-0870(A),
   087.C-0831(A),
   and 183.C-0972(A).
   }\fnmsep
   \thanks{
   Tables~\ref{parameters_table_full},
   ~\ref{abundance_table_full}, and ~\ref{carbon_oxygen_MA15_stars} are only available in the electronic version of the paper or
   at the CDS via anonymous ftp to cdsarc.u-strasbg.fr (130.79.128.5)
   or via http://cdsweb.u-strasbg.fr/cgi-bin/qcat?J/A+A/}
}


    \author{J. Maldonado
          \inst{1}
          \and
          E. Villaver
          \inst{2}
          }

          \institute{INAF - Osservatorio Astronomico di Palermo,
                     Piazza Parlamento 1, I-90134 Palermo, Italy
          \and
           Universidad Aut\'onoma de Madrid, Dpto. F\'isica Te\'orica, M\'odulo 15,
           Facultad de Ciencias, Campus de Cantoblanco, 28049 Madrid, Spain
            }

   \offprints{J. Maldonado \\ \email{jmaldonado@astropa.inaf.it}}
   \date{Received ; accepted }

 
  \abstract
   { Detailed chemical abundance studies have revealed
    different trends between samples of planet and non-planet hosts.
    Whether these trends are related to the presence of planets or not is strongly
    debated. At the same time, tentative evidence that the properties of evolved stars with planets
    may be different from what we know for main-sequence hosts has been recently
    reported.
    }
   {We aim to test whether evolved stars with planets show any chemical peculiarity that
    could be related to the planet formation process.}
   {
    We determine in a consistent way the metallicity and individual abundances of a
    large sample of evolved (subgiants and red giants) and main-sequence stars with and without known planetary companions,
    and discuss their metallicity distribution and trends.
    Our methodology is based on the analysis
    of high-resolution \'echelle spectra (R $\gtrsim$ 57 000) from 2-3 m class telescopes.
    It includes the calculation of the fundamental stellar
    parameters, as well as, individual abundances
    of C, O , Na, Mg, Al, Si, S, Ca, Sc, Ti, V, Cr, Mn, Co, Ni, and Zn.     
   }
   {
    No differences  in the
    <[X/Fe]> vs. condensation temperature (T$_{\rm C}$)
    slopes are found between the samples
    of planet and non-planet hosts when all elements are considered. However, if the analysis
    is restricted to only refractory elements, differences in the T$_{\rm C}$-slopes  between
    stars with and without known planets are found. This result is found to be dependent on
    the stellar evolutionary stage, as it holds for main-sequence and subgiant stars, while
    there seem to be no difference between planet and non-planet hosts among the sample of giants.
    A search for correlations between the T$_{\rm C}$-slope and the stellar properties
    reveals significant correlations with the stellar mass and the stellar age.
    The data also suggest that differences 
    in terms of mass and age
    between main-sequence planet and non-planet hosts may be present.
}
   { Our results are well explained by radial mixing in the Galaxy. The sample of giant
     contains stars more massive and younger than their main-sequence
     counterparts. This leads to a sample of stars possibly less contaminated by stars not born in the solar
     neighbourhood, leading to no chemical differences between planet and 
     non planet hosts. 
     The sample of main-sequence stars may contain more
     stars from the
     outer disc  (specially the non-planet host sample) 
     which might led to the differences observed in the chemical trends.
}


   \maketitle

\section{Introduction}\label{introduccion}

  Detailed chemical analysis of large samples of stars hosting planets 
  have revealed as a powerful technique to help in our understanding
  of how planetary systems do form and evolve.
  However, besides the increasing number of recent studies
  \citep[e.g.][]{2015A&A...583A.135B,2015A&A...580A..24D,
  2015A&A...579A..20M,2015A&A...579A..52N,
  2015ApJ...808...13R,2015A&A...580A..30T},
  the only well established correlation found so far
  is the one
  that relates the stellar metallicity with the probability of hosting a gas giant planet
  \citep[e.g.][]{1997MNRAS.285..403G,2004A&A...415.1153S,2005ApJ...622.1102F}.
  Any other claim of a chemical trend in planet-hosting stars has so far been disputed.

  \cite{2009ApJ...704L..66M} reports a deficit of refractory
  elements in the Sun with respect to other solar twins concluding that
  it is related to the formation of terrestrial planets.
  Similar chemical patterns are found by 
  \cite{2009A&A...508L..17R,2010A&A...521A..33R} and \cite{2011MNRAS.416L..80G}
  in other solar twins and analogs. 
  This interpretation has however been challenged by other works that point towards
  galactic chemical evolution effects \citep{2010ApJ...720.1592G,2013A&A...552A...6G}
  or towards an age/inner galactic origin of the planet hosts stars
  as the cause of the detected small chemical depletions
  \citep{2014A&A...564L..15A}. 
  In a recent work, \citet[][hereafter MA15]{2015A&A...579A..20M} reports chemical
  different slopes in the abundance versus elemental condensation temperature diagram
  between stars with cool gas-giant planets
  and non-planet hosts, noting as well moderate correlations between
  the abundance-condensation temperature trend and stellar properties such as
  age or metallicity. 

  While most of the detailed chemical studies done so far are around 
  main-sequence (MS) stars, little is known about possible chemical trends in evolved stars
  with planets.
  For instance, it is still unclear whether giant stars with planets
  follow the gas-giant planet stellar metallicity correlation
  \citep{2005PASJ...57..127S,2005ApJ...632L.131S,2007A&A...475.1003H,
  2007A&A...473..979P,2008PASJ...60..781T,2010ApJ...725..721G}.
  With the wealth of new planetary discoveries in the last years we re-visit
  this issue by performing homogeneous observations and analysis of a
  large sample of 142 evolved stars.  
  In
  \citet[][hereafter MA13]{2013A&A...554A..84M},
  we find that whilst the metallicity distribution of planet-hosting giant stars with stellar masses
  M$_{\star}$ $>$ 1.5 M$_{\odot}$ follow the
  general trend established for the main sequence stars hosting planets,
  giant planet hosts in the mass domain M$_{\star}$ $\le$ 1.5 M$_{\odot}$
  do not show metal enrichment.
  Similar results are found by \cite{2013A&A...557A..70M}. 
  However, \cite{2015A&A...574A..50J} 
  does not find any clear metallicity difference
  between giant stars as planet hosts and non hosts for M$_{\star}$ $>$ 1.5 M$_{\odot}$.
  \cite{2015A&A...574A.116R} explore the planet occurrence
  rate with stellar metallicity and stellar mass
  (exploring the mass range 1.0-3.8 M$_{\odot}$)
  in the UCO/Lick survey.
  The authors perform a distinction between ``secure'' (15 stars) planet hosts and
  planet ``candidates'' (20 stars)  based on their available data and found a strong planet-metallicity
  correlation among the secure planet hosts,
  and attribute the lack of correlation found on the sample of planet candidates
  to the fact that the candidate planets are found preferentially among 
   stars with rather small metallicity and mass.
  The fact that the bulk of their candidate planets is found
  among their less massive and low-metallicity stars
  is intriguing at least.

  Further, the detection of planetary companions is hampered 
  by the large levels of stellar jitter in evolved stars introduced
  by stellar p-mode oscillations, which may reach $\sim$ 100 m s$^{-1}$ 
  \citep{1995A&A...293...87K}. 
  \cite{2016A&A...585A..73N} show that the minimum detectable planetary mass
  is a increasing function of the orbital separation and the
  stellar luminosity, being the detection of close-in, small planets
  (M$_{\rm p}\sin i$ $<$ 2 M$_{\rm J}$ within 1 au) a difficult task
  when dealing with evolved stars.

  In this paper a detailed analysis of the chemical abundances of
  a large sample of evolved (subgiants and red giants) stars
  with and without planets is presented. We aim to test
  whether these stars show any chemical peculiarity, and to unravel
  their origin.
  This work follow the analysis presented in
  MA13, but now
  we extend it to studying possible trends between the abundances
  and the elemental condensation temperature. 
  This paper is organised as follows. Sect.~\ref{secction_observations}
  describes the stellar samples analysed in this work, the spectroscopic
  data in which the work is based and the analysis.
  The distribution of abundances are presented in Sect.~\ref{section_analysis}.
  The results are discussed in Sect.~\ref{seccion_discussion}.
  Our conclusions follow in Sect.~\ref{conclusions}.

\section{Data and spectroscopic analysis}\label{secction_observations}
\subsection{Stellar sample}

\begin{figure}
\centering
\includegraphics[angle=270,scale=0.45]{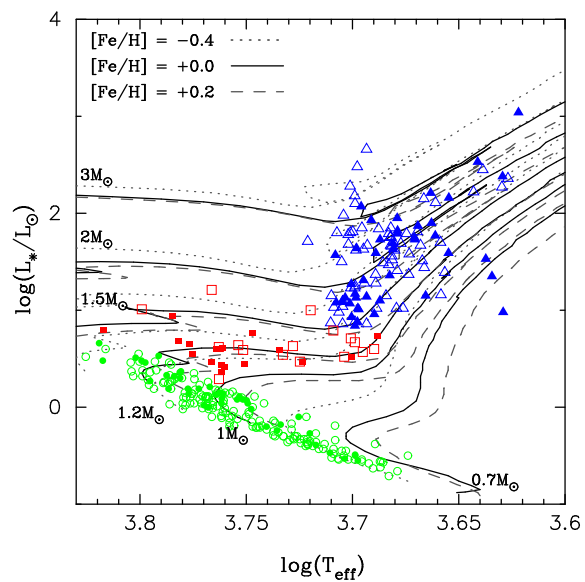}
\caption{
Luminosity versus T$_{\rm eff}$ diagram for the observed stars.
Giants are plotted with blue triangles,
subgiants with red squares, and MS stars with green circles. Filled symbols
indicate planet hosts. Some evolutionary tracks ranging from 0.7 to 3.0 solar masses
from \cite{2000A&AS..141..371G}
are overplotted. For each mass, three tracks are plotted, corresponding to
Z=0.008 ([Fe/H]=-0.4 dex, dotted lines), Z=0.019 ([Fe/H]=+0.0 dex, solid lines),
and Z=0.030 ([Fe/H]=+0.20 dex, dashed lines).
}
\label{diagrama_hr}
\end{figure}

  Figure~\ref{diagrama_hr} shows the Hertzsprung-Russell (HR) diagram of
  the observed stars. The total number of stars amounts up to 341.
  They are classified as red giants (blue triangles,
  giants from now on), subgiants (red squares), and main-sequence stars
  (green circles).
  The samples of giant and subgiant stars were built using as reference
  the stars listed in MA13, with additional data for twelve new subgiants.
  The list of MS stars comes from MA13 and M15 works
  and have been analysed homogeneously.
  According to their luminosity class and taking into account the
  presence (or absence) of planetary companions
  \footnote{According to the available data at the Extrasolar Planets
  Encyclopaedia, http://exoplanet.eu/},
  our sample is divided
  into 43 giant stars with known planets (hereafter GWPs),
  67 giant stars without planets (GWOPs), 16 subgiants hosting planets (SGWPs),
  17 subgiants without planets (SGWOPs),
  41 MS stars harbouring planets (MSWPs), and 157 MS stars without known
  planets (MSWOPs). 
  We note that the total number of giant stars known to host planets is 68
  so our GWP sample is statistically representative although it does
  not include the stellar hosts non-observable from the northern hemisphere.


\begin{figure}
\centering
\begin{minipage}{0.49\linewidth}
\includegraphics[angle=270,scale=0.45]{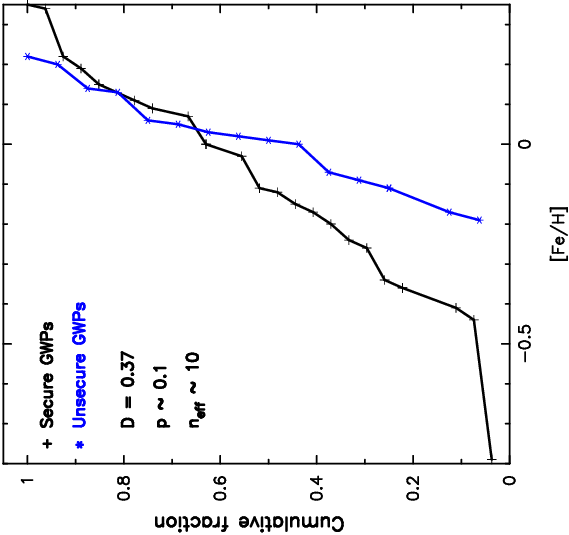}
\end{minipage}
\begin{minipage}{0.49\linewidth}
\includegraphics[angle=270,scale=0.45]{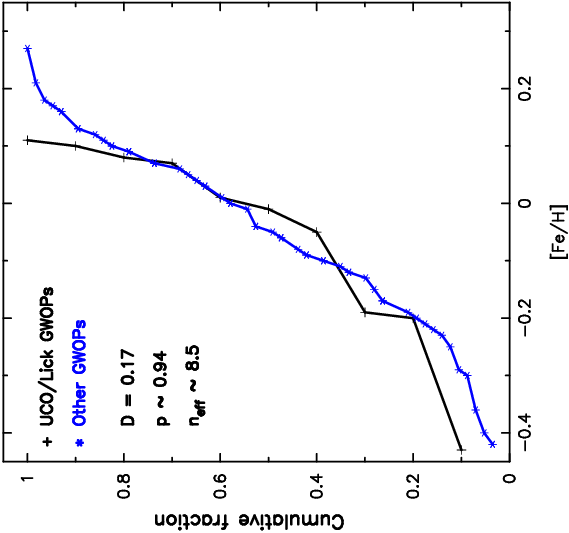}
\end{minipage}
\caption{
  Cumulative [Fe/H] distribution for giant stars. Left: GWPs, divided into ``secure''
   (27 stars) and ``unsecure''  (16 stars) planet hosts. Right: GWOPs, divided into stars included
  in the UCO/Lick survey (10 stars)  and GWOPs non-included in this survey  (57 stars). See text for details.
}  
\label{metal_plots_heidelberg}
\end{figure}

  Before continuing with the analysis we have first checked for the presence of
  biases in our sample that might influence the results. 
  Recently, \cite{2015A&A...574A.116R}
  suggested that previous samples of GWPs may be contaminated
  by ``unsecure'' planet detections, mainly around low-metallicity
  and low-mass stars
  since this seems to be the case in the sample they analysed.
  In order to test whether this could be 
  our case, 
  we have 
  checked whether 
  our original sample of stars with planets 
  overlap with the list of planets included in the
  web-page maintained by these authors\footnote{http://www.lsw.uni-heidelberg.de/users/sreffert/giantplanets.html}. 
  Note that those are radial velocity planets published
  in the literature and therefore each data set is subject to different
  selection criteria than that applied to differentiate among secure/candidate planet candidates 
  in the sample of \cite{2015A&A...574A.116R}. 
  Obviously, the criteria used by those authors to include (or not)
  a planet detection in their web-page may be discussed, but this is out of the scope
  of this work. Here, we just assumed their criteria is valid and check for biases in our sample. 
  From our original list of 43 GWPs, 27 overlap with the aforementioned
  list, while 16 are not included in this list.   
  The analysis of the metallicity distribution of secure/unsecure
  planets hosts, Figure~\ref{metal_plots_heidelberg} (left),
  reveals that stars with ``unsecure'' planets
  do not show lower metallicities, but rather higher metallicities
  than the stars with secure planets. While the median metallicity
  of the secure planets hosts is -0.11 dex, ``unsecure'' planets hosts
  show a median value of +0.02 dex. Further, a two-sided Kolmogorov-Smirnov
  (KS) test returns a probability of $\sim$ 0.1 of both distributions
  being drawn from the same parent population. 
  Therefore, we conclude that the metallicity distribution of the 16 planets not included
  in the Reffert's list do  not seem to contaminate our sample
  towards low-metallicity stars.

  Our GWOP sample was selected based on available giant stars from the
  \cite{2008AJ....135..209M} list of {\sc Hipparcos} giants within
  100 pc from the Sun. So, in principle
  we cannot rule out the possibility
  that some of these stars host an undetected gas-giant planet. 
  \cite{2015A&A...574A.116R} claim planet detection of 4-5\% in their sample.
  A slightly higher detection rate of 10-15\% is provided by the Th\"uringer Landessternwarte
  survey \citep{2009A&A...505.1311D}.
  \cite{2011ApJS..197...26J} analyse 246 subgiants from the California Planet Search,
  providing a detection rate of 15\%.
  These numbers are consistent with the models by \cite{2008ApJ...673..502K}
  which predict a frequency of gas-giant planets of 10\% around 1.5 $M_{\odot}$ stars.
  To test whether our GWOP sample may be contaminated, we divided it into two subsamples:
  one including those stars which have been monitored in the UCO/Lick survey 
  as listed in \cite{2015A&A...574A.116R} and have not
  been reported to have a detected planet (10 stars);
  and another subsample with our GWOP stars not included in this survey
  (57 stars).
  Figure~\ref{metal_plots_heidelberg} (right), shows the  
  metallicity distribution of both subsamples. 
  From the plot is clear that they are almost identical. A KS test
  provides a probability of 94\% of both subsamples having similar metallicity
  distributions. 
  Thus, we conclude that it is very unlikely that the properties of our GWOP 
  (such as metallicity or elemental abundances)
  are affected in a significant way by the presence of undetected 
  gas-giant planets.
  This result seems in line with the contamination expected in the GWOP sample
  (as seen, at most at the 10\% level), too 
  small to significantly affect the results /shift the metallicity distributions.

  Considering less massive planets
  (M$_{\rm p}\sin i$ $<$ 30 M$_{\oplus}$),
  there is increasing evidence that they
  might be common around MS solar-type stars
  \citep[e.g.][]{2011arXiv1109.2497M,2012Natur.481..167C,2013Natur.503..381H},
  which may certainly contaminate the MSWOP sample.
   We note that stars hosting low-mass planets
  do not seem to be preferentially metal rich
  \citep[][although see \cite{2015AJ....149...14W}, for an opposing view]{2010ApJ...720.1290G,2011arXiv1109.2497M,2011A&A...533A.141S,2012Natur.486..375B,2015ApJ...808..187B}.
  Furthermore, MA15 find that stars with low-mass planets show similar
  chemical trends that stars without known planetary companions.
   Therefore we discarded from our MSWP sample those stars
  harbouring low-mass planets.

  Regarding giant stars,
  the detection of low mass planets is hindered around this kind of stars 
  \citep[see e.g.][]{2016A&A...585A..73N}.
  Therefore any hint of giant stars hosting more massive planets than MS stars (e.g. MA13)
  needs to be interpreted with caution as  the larger levels of stellar oscillation in evolved
  stars certainly introduce an observational bias against the detection of low-mass planets via
  the radial velocity method.
  In addition, it is relevant to note that low-mass planets
  have more chances of surviving the  processes that take place when
  a MS star evolves off  the MS as there is a strong dependence of the tidal forces
  on the mass ratio of the planet-star system  
  \citep[e.g.][]{2009ApJ...705L..81V,2012ApJ...761..121M,2014ApJ...794....3V}.

   We finally note that the mass distribution
   of our giant stars is similar to those of more general exoplanet
   search projects focused on evolved stars
   (see next subsection for details on this calculation).
   Our evolved stars sample covers a mass range between
   1.0 and 3.6 M$_{\odot}$ having
   a peak at $\sim$ 1.5 M$_{\odot}$ and then decreases steadily,
   as does the mass distribution of the giant stars of
   the  PTPS survey or the  Retired A
   Stars project  \citep[see][Figure 13c]{2016A&A...585A..73N}.
      
\begin{figure}
\centering
\includegraphics[angle=270,scale=0.45]{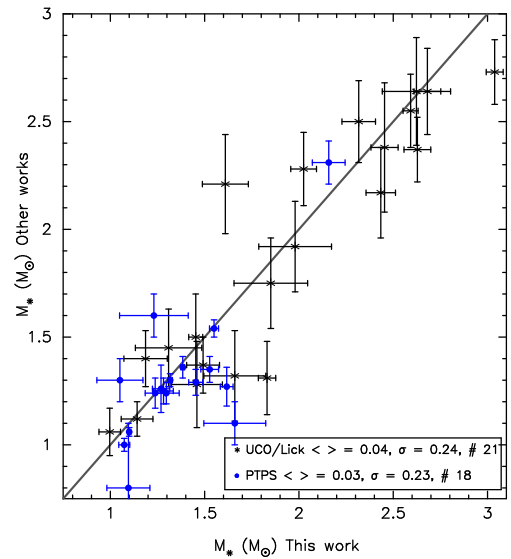}
\caption{
Stellar mass values from the literature estimates versus the values obtained in this work.
The symbol $< >$
in the legend represents the median difference. The Gran continuous
line represents the 1:1 relation.
}
\label{comparison_masas}
\end{figure}

   Recent works have put into question the reliability of the masses
   of evolved stars hosting planetary systems 
   \citep{2011ApJ...739L..49L,2013ApJ...774L...2L,2013ApJ...772..143S}.
   Therefore, a comparison between our derived masses and those
   given in the  PTPS \citep{2012A&A...547A..91Z,2016A&A...585A..73N}
   and in the UCO/Lick \citep{2015A&A...574A.116R} surveys were performed.
   The masses provided in other surveys  might be overestimated
   \citep[see][]{2016A&A...585A..73N}   
   and thus we do not use them for comparison with our sample. 
   The Retired A stars and their Companions survey \citep{2015ApJ...812...96G}
   also provides masses consistent with the PTPS and Lick surveys,
   however, we do not have enough stars in common to include them
   in the analysis. 
   The comparison is shown in Figure~\ref{comparison_masas}.
   It reveals an overall good agreement between
   our mass estimates and those by PTPS and Lick surveys.
   Note that the median difference is only
   0.03, 0.04 solar masses with a rms standard deviation of 
   0.23, and 0.24 M$_{\odot}$ for the PTPS and Lick
   surveys respectively. 


\subsection{Spectroscopic analysis}

  The spectroscopic data used in this work is basically the same used
  in MA13 and MA15 to which we refer to for further details.  
  In brief, high-resolution spectra of the stars were obtained:
  i) at La Palma observatory (Canary Islands, Spain) with the HERMES
  spectrograph \citep{2011A&A...526A..69R} at the MERCATOR telescope;
  ii) at the Nordic Optical Telescope with the FIES \citep{1999anot.conf...71F}
  instrument;
  iii) at the  2.2-metre telescope of the Calar Alto
  observatory (CAHA, Almer\'ia, Spain) using the the FOCES \citep{foces} spectrograph;
  iv) at the Telescopio Nazionale Galileo (TNG, 3.58 m) using the 
  SARG \citep{sarg} spectrograph.
  Additional spectra  from  the  public
  library  ``S$^{4}$N'' \citep{s4n} as well as HARPS and FEROS
  spectra from the ESO
  Archive Facility\footnote{http://archive.eso.org/wdb/wdb/adp/phase3\_main/form}
  were also used.

   We are aware that ideally, all our targets should
   have been observed with the same spectrograph and configuration.
   However, all the spectra used in this work have high resolution
   (from $\sim$ 42000 of FEROS spectra to $\sim$ 115000 for HARPS),
   have a high signal-to-noise ratio (median value 107 at 6050 \AA) and cover a wide
   spectral range (from 3780-6910 \AA \space for HARPS to
   3400-10900 \AA \space for McDonald)
   with enough lines for a high-quality abundance
   determination.  


  Basic stellar parameters T$_{\rm eff}$, $\log g$, microturbulent
  velocity $\xi_{\rm t}$, and [Fe/H] are determined using the code
  {\sc TGVIT}\footnote{http://optik2.mtk.nao.ac.jp/\textasciitilde{}takeda/tgv/}
  \citep{2005PASJ...57...27T}, which implements the iron ionisation
  and excitation equilibrium  conditions, a methodology
  widely applied to solar-like stars. 
  The line list as well as the adopted parameters (excitation potential,
  $\log (gf)$ values) can be found on Y. Takeda's web page. This code
  makes use of ATLAS9, plane-parallel, LTE atmosphere models
  \citep{1993KurCD..13.....K}. 

  Chemical abundance of individual elements
  C, O, Na, Mg, Al, Si, S, Ca, Sc, Ti, V, Cr,  Mn, Co, Ni,
  and Zn
  were obtained using the 2014 version of the
  code {\sc MOOG}\footnote{http://www.as.utexas.edu/\textasciitilde{}chris/moog.html}
  \citep{1973PhDT.......180S} together with ATLAS9 atmosphere models
  \citep{1993KurCD..13.....K}. The measured equivalent widths (EWs) of a list of narrow,
  non-blended lines for each of the aforementioned species are used as inputs. The selected
  lines are taken from the list provided by MA15. 
  Hyperfine structure (HFS) was taken into account for
  V~{\sc i}, and Co~{\sc i} abundances.
   HFS corrections for Mn~{\sc i} were not taken into account
  as in MA15 we found slightly different abundances when considering
  different lines.
  Although HFS effects may be present for other elements
  (e.g. Mg~{\sc i}, Sc~{\sc i}),
  we do not expect these
  effects to bias the results of the comparisons performed in this
  work between samples of stars with and without
  planets, given that they have otherwise similar properties.

  The oxygen abundance was derived from the forbidden [O~{\sc i}] line at 
  6300 \space \AA. This line is well known to be blended with a closer Ni~{\sc i} line
  \citep[e.g.][]{2001ApJ...556L..63A}.
  We made use of the {\sc MOOG} driver {\it ewfind} to determine the
  EW of the Ni line using the previously derived  Ni
  abundance. This EW was subtracted from the measured EW of the
  of the Ni{~\sc i} plus [O~{\sc i}] feature. Then, the oxygen 
  abundance was determined from the remaining EW
  \citep[e.g.][]{2010ApJ...725.2349D,2013A&A...552A...6G}. 
  Since oxygen abundances in MA15 stars were derived from
  the O~{\sc i} triplet lines at 777 nm, we have recomputed them 
  using the [O~{\sc i}]  630 nm line. 
  For the carbon abundance the lines at 505.2 and 538.0 nm were used
  instead of those reported in MA15, since we found the latest to
  give abnormally high abundances for the giant stars.
  Some problems when using the 505.2 nm line have been reported by
  \cite{2011A&A...526A..71D}.
  However, we do not find significant differences between the abundances
  derived from the 505.2 nm line and the ones derived from the 538.0 nm
  line, being the mean and median differences around $\sim$ 0.02. 

  Evolutionary parameters namely, stellar mass, radius, and age
  were computed using the code {\sc Param}\footnote{http://stev.oapd.inaf.it/cgi-bin/param}
  \citep{2006A&A...458..609D} with the new PARSEC isochrones from 
  \cite{2012MNRAS.427..127B}. 

  Our derived stellar parameters are given in Table~\ref{parameters_table_full},
  whilst the abundances are provided in Table~\ref{abundance_table_full}.
  The recomputed abundances of carbon and oxygen for those stars taken from MA15
  are given in Table~\ref{carbon_oxygen_MA15_stars}.
  These tables are available at the CDS. 

\onllongtab{
\begin{longtable}{llcccccccc}
\caption{
 Spectroscopic parameters with uncertainties
 for the stars measured in this work. 
  }\label{parameters_table_full}\\
\hline
\hline
HIP & HD & T$_{\rm eff}$ &  $\log g$            & $\xi_{t}$           & [Fe/H] & Sp.$^{\dag}$   &  Age  & M$_{\star}$   & R$_{\star}$    \\
    &    &  (K)          & (${\rm cm s^{-2}}$)  & (${\rm km s^{-1}}$) & dex    &                & (Gyr) & (M$_{\odot}$) & (R$_{\odot}$)  \\
 (1)& (2)&  (3)          & (4)                  & (5)                 & (6)    & (7)            & (8)   & (9)           & (10)           \\
\hline
\endfirsthead
\caption{Continued.} \\
\hline
HIP & HD & T$_{\rm eff}$ &  $\log g$            & $\xi_{t}$           & [Fe/H] & Sp.$^{\dag}$   &  Age  & M$_{\star}$   & R$_{\star}$    \\
    &    &  (K)          & (${\rm cm s^{-2}}$)  & (${\rm km s^{-1}}$) & dex    &                & (Gyr) & (M$_{\odot}$) & (R$_{\odot}$)  \\
 (1)& (2)&  (3)          & (4)                  & (5)                 & (6)    & (7)            & (8)   & (9)           & (10)           \\
\hline
\endhead
\hline
\endfoot
\hline
\endlastfoot
\hline\noalign{\smallskip}
\multicolumn{10}{c}{Giant stars with planets}\\
\hline\noalign{\smallskip}
1692	&	1690	&	4338	$\pm$	20	&	2.04	$\pm$	0.08	&	1.56	$\pm$	0.14	&	-0.24	$\pm$	0.04	&	3	&	6.80	$\pm$	3.12	&	1.11	$\pm$	0.15	&	18.60	$\pm$	2.72	\\
4297	&	5319	&	4895	$\pm$	28	&	3.33	$\pm$	0.09	&	1.10	$\pm$	0.10	&	0.05	$\pm$	0.03	&	3	&	4.48	$\pm$	0.85	&	1.30	$\pm$	0.07	&	3.76	$\pm$	0.35	\\
10085	&	13189	&	4187	$\pm$	35	&	1.60	$\pm$	0.14	&	1.49	$\pm$	0.17	&	-0.36	$\pm$	0.06	&	4	&	4.41	$\pm$	2.89	&	1.24	$\pm$	0.27	&	32.32	$\pm$	6.08	\\
12247	&	16400	&	4864	$\pm$	23	&	2.65	$\pm$	0.08	&	1.41	$\pm$	0.10	&	-0.03	$\pm$	0.03	&	3	&	1.68	$\pm$	0.18	&	1.83	$\pm$	0.09	&	9.79	$\pm$	0.36	\\
20889	&	28305	&	4911	$\pm$	25	&	2.73	$\pm$	0.09	&	1.39	$\pm$	0.13	&	0.15	$\pm$	0.04	&	4	&	0.63	$\pm$	0.07	&	2.63	$\pm$	0.07	&	12.45	$\pm$	0.26	\\
36616	&	59686	&	4669	$\pm$	20	&	2.63	$\pm$	0.07	&	1.32	$\pm$	0.11	&	0.13	$\pm$	0.04	&	4	&	1.48	$\pm$	0.46	&	1.98	$\pm$	0.19	&	12.57	$\pm$	0.44	\\
37826	&	62509	&	4887	$\pm$	18	&	3.00	$\pm$	0.06	&	1.23	$\pm$	0.10	&	0.09	$\pm$	0.03	&	4	&	1.31	$\pm$	0.17	&	2.03	$\pm$	0.07	&	8.63	$\pm$	0.16	\\
\end{longtable}
\tablefoot{$^{\dag}$Spectrograph: {\bf(1)} CAHA/FOCES; {\bf(2)} TNG/SARG; {\bf(3)} NOT/FIES; {\bf(4)} MERCATOR/HERMES; {\bf(5)} S$^{4}$N-McD;
{\bf(6)} S$^{4}$N-FEROS; {\bf(7)} ESO/FEROS; {\bf(8)} ESO/HARPS}
}


\onllongtab{
\begin{longtab}
\begin{landscape}
\begin{longtable}{lrrrrrrrrrrrrrrrrrrr}
\caption{\label{abundance_table_full} Derived abundances [X/H]}\\
\hline\hline
 Star    &          C~{\sc i}  & O~{\sc i}  & Na~{\sc i} & Mg~{\sc i} & Al~{\sc i} &
                   Si~{\sc i} & S~{\sc i}  & Ca~{\sc i} & Sc~{\sc i} & Sc~{\sc ii} &
                   Ti~{\sc i} & Ti~{\sc ii} & V~{\sc i} & Cr~{\sc i} & Cr~{\sc ii} &
                   Mn~{\sc i} & Co~{\sc i}  & Ni~{\sc i} & Zn~{\sc i}\\
\hline
\endfirsthead
\caption{continued.}\\
\hline\hline
 Star   &          C~{\sc i}  & O~{\sc i}  & Na~{\sc i} & Mg~{\sc i} & Al~{\sc i} &
                   Si~{\sc i} & S~{\sc i}  & Ca~{\sc i} & Sc~{\sc i} & Sc~{\sc ii} &
                   Ti~{\sc i} & Ti~{\sc ii} & V~{\sc i} & Cr~{\sc i} & Cr~{\sc ii} &
                   Mn~{\sc i} & Co~{\sc i}  & Ni~{\sc i} & Zn~{\sc i}\\
\hline
\endhead
\hline
\endfoot
\hline\noalign{\smallskip}
\multicolumn{20}{c}{Giant stars with planets}\\
\hline\noalign{\smallskip}
HIP1692 &        0.13 &    -0.06 &   -0.15 &   -0.09 &   0.06 &    -0.03 &   0.53 &    -0.36 &   -0.01 &   -0.15 &   -0.04 &   -0.14 &   0.20 &    -0.27 &   -0.18 &   0.07 &    0.01 &    -0.19 &   -0.28 \\ 
              &  0.05 &           0.03 &           0.09 &           0.05 &           0.04 &           0.06 &           0.22 &           0.14 &           0.33 &           0.06 &           0.09 &           0.11 &           0.13 &           0.07 &           0.08 &           0.09 &           0.09 &           0.04 &           0.23 \\         
HIP4297 &        0.02 &    0.02 &    0.13 &    0.17 &    0.20 &    0.12 &    0.30 &    0.01 &    0.04 &    0.01 &    0.04 &    0.07 &    0.09 &    0.02 &    0.02 &    0.37 &    0.08 &    0.06 &    0.06 \\  
              &  0.07 &           0.04 &           0.04 &           0.05 &           0.02 &           0.04 &           0.07 &           0.07 &           0.17 &           0.06 &           0.07 &           0.10 &           0.06 &           0.06 &           0.08 &           0.10 &           0.06 &           0.04 &           0.15 \\         
HIP10085 &       0.16 &    -0.10 &   -0.15 &   -0.02 &   -0.10 &   -0.03 &   0.58 &    -0.37 &   -0.18 &   -0.28 &   -0.18 &   -0.12 &   0.11 &    -0.31 &   -0.37 &   -0.07 &   -0.17 &   -0.28 &   -0.35 \\ 
              &  0.08 &           0.03 &           0.07 &           0.05 &           0.09 &           0.07 &           0.05 &           0.12 &           0.46 &           0.06 &           0.08 &           0.13 &           0.10 &           0.06 &           0.09 &           0.11 &           0.07 &           0.04 &           0.05 \\         
HIP12247 &       -0.26 &   -0.04 &   0.11 &    0.20 &    0.07 &    0.09 &    0.20 &    -0.02 &   -0.13 &   -0.07 &   -0.02 &   0.04 &    -0.01 &   -0.04 &   -0.10 &   0.28 &    -0.02 &   -0.07 &   -0.03 \\
\end{longtable}
\tablefoot{$^{\dag}$: GSC02883-01687}
\end{landscape}
\end{longtab}
}

\onllongtab{
\begin{longtable}{lcc}
\caption{
 Carbon and oxygen abundances for the MS stars from MA15.
  }\label{carbon_oxygen_MA15_stars}\\
\hline
\hline
 HIP    &          C~{\sc i}  & O~{\sc i} \\
\hline
\endfirsthead
\caption{Continued.} \\
\hline
 HIP    &          C~{\sc i}  & O~{\sc i} \\
\hline
\endhead
\hline
\endfoot
\hline
\endlastfoot
\hline\noalign{\smallskip}
171	&	-0.52	$\pm$	0.02	&	-0.25	$\pm$	0.02	\\
490	&	-0.14	$\pm$	0.05	&				\\
522	&	0.04	$\pm$	0.03	&				\\
544	&	0.00	$\pm$	0.02	&				\\
1598	&	-0.30	$\pm$	0.01	&				\\
1599	&	-0.26	$\pm$	0.02	&	-0.09	$\pm$	0.01	\\
2941	&				&	0.15	$\pm$	0.02	\\
3093	&	0.09	$\pm$	0.04	&	0.13	$\pm$	0.02	\\
\end{longtable}
}

\section{Analysis}\label{section_analysis}
\subsection{[X/Fe]-T$_{\rm C}$ trends}
\label{xfe_tc_trends}

 Chemical differences were searched for by studying possible trends
 between the abundances, [X/Fe], and the elemental condensation temperature,
 T$_{\rm C}$.
 Mean abundances for each of the samples 
 were computed, and the T$_{\rm C}$-slope was derived
 by performing a linear fit, weighting each element by its
 corresponding star-to-star scatter.
 Values of T$_{\rm C}$ correspond
 to a 50\% equilibrium condensation temperature for a solar system composition gas
 \citep{2003ApJ...591.1220L}. 

\begin{figure*}
\centering
\begin{minipage}{0.49\linewidth}
\includegraphics[angle=270,scale=0.375]{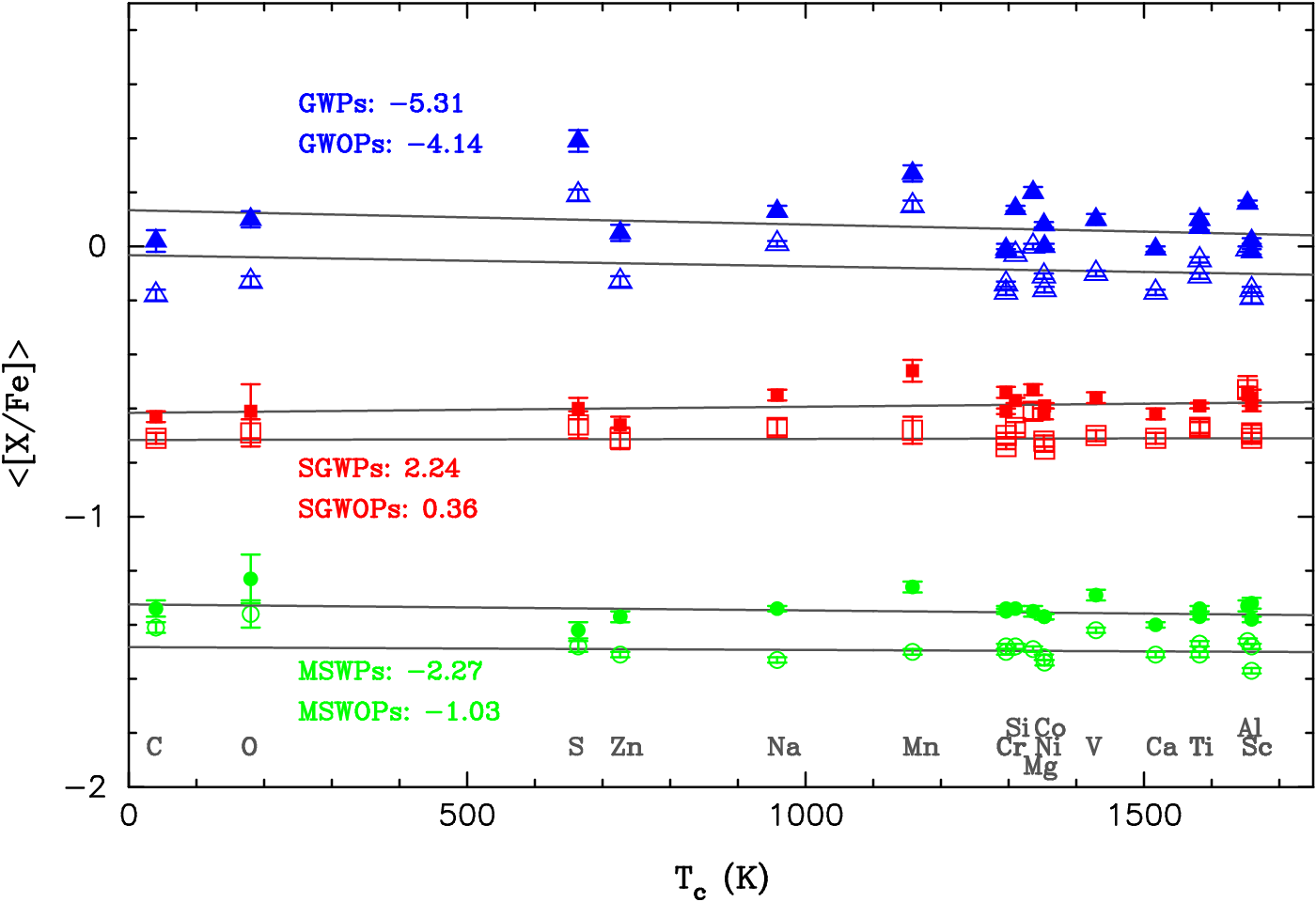}
\end{minipage}
\begin{minipage}{0.49\linewidth}
\includegraphics[angle=270,scale=0.375]{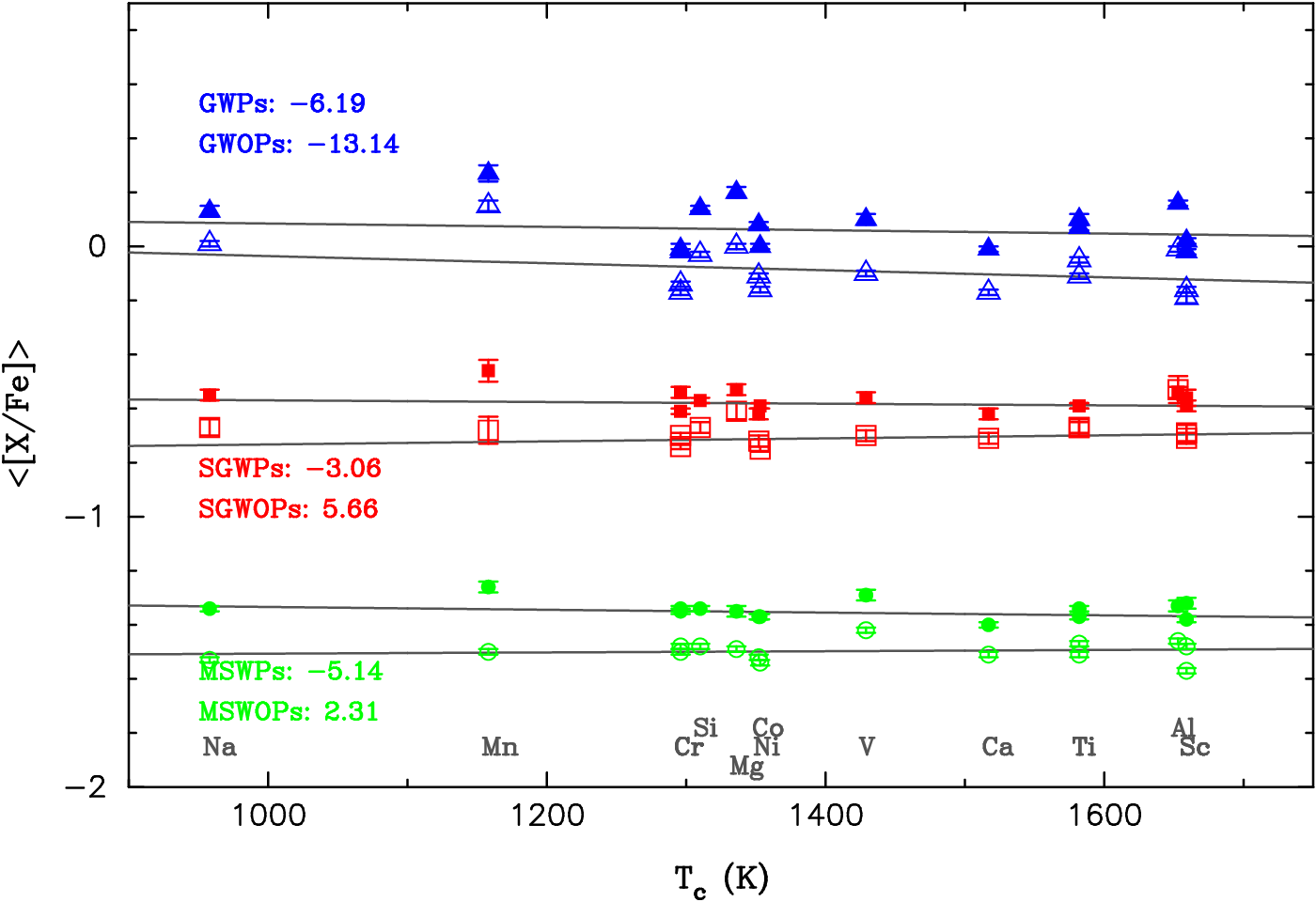}
\end{minipage}
\caption{
  <[X/Fe]>-T$_{\rm C}$ trends of all the stars analysed. Giants are
 plotted with blue triangles, subgiants with red squares, and MS stars
 with green circles. Filled symbols indicate planet hosts.
 Each planet host subsample is shown against its corresponding comparison subsample
 (e.g. GWPs vs. GWOPs)
 with an offset of -0.15 for the sake of clarity.
 The offset between giants, subgiants, and MS samples is -0.75.
 For guidance, the derived slopes are shown
 in the plots (units of 10$^{\rm -5}$ dex/K).
 The left panel shows the   <[X/Fe]>-T$_{\rm C}$ trends
 when all elements
 (volatiles and refractories) are taken into account whilst the
 right one shows the <[X/Fe]>-T$_{\rm C}$ trends when only refractories are considered.
}
\label{tc_plots_giants}
\end{figure*}

 As in MA15 we compute the slope of the 
 [X/Fe] vs. T$_{\rm C}$ fit considering firstly all refractory and volatile
 elements (T$^{\rm all}_{\rm C}$-slope), and then considering only refractories
 (T$^{\rm refrac}_{\rm C}$-slope). In this way we take into account the fact
 that the abundances of volatiles are in general more difficult to obtain accurately 
 \footnote{We consider as volatile those elements
 with T$_{\rm C}$ lower than 900 K, namely C, O, S, and Zn.}.
 To give a significance for the derived slopes a Monte Carlo simulation was carried out.
 We created 10$^{4}$ series of simulated
 random abundances and errors,  keeping the media and the standard deviation of the original data.
 For each series of simulated data the corresponding  T$_{\rm C}$-slope was derived. Assuming
 that the distribution of the simulated slopes follows a Gaussian function we then compute
 the probability that the simulated slope takes the value found when fitting the original data
 (hereafter $p$-value). 
 The corresponding plots are shown in Figure~\ref{tc_plots_giants}, and a summary
 of the fits is presented in Table~\ref{linearfits}.

\begin{table}
\centering
\caption{Results of the  <[X/Fe]>-T$_{\rm C}$ linear fits. For each fit its
 probability of slope ``being by chance'' ($p$) is also given.}
\label{linearfits}
\begin{scriptsize}
\begin{tabular}{lcccc}
\hline\noalign{\smallskip}
Sample   & \multicolumn{2}{c}{\bf All elements}               & \multicolumn{2}{c}{\bf Only refractory}        \\
         & slope ($\times$10$^{\rm -5}$dex/K)       & $p$.    & slope ($\times$10$^{\rm -5}$dex/K)  &  $p$     \\
\hline
 GWPs    &  -5.31 $\pm$ 1.20   &   0.06  &         -6.19  $\pm$       1.84  &         0.07   \\
 GWOPs   &  -4.14 $\pm$ 0.81   &   0.07  &        -13.14  $\pm$       1.45  &         0.05   \\
\hline
 SGWPs   & 2.24 $\pm$   1.17   &   0.12  &         -3.06  $\pm$       2.32  &         0.17   \\
 SGWOPs  & 0.36 $\pm$   1.29   &   0.17  &          5.66  $\pm$       3.35  &         0.17   \\
\hline
 MSWPs   & -2.27$\pm$   1.11  &    0.11  &         -5.14  $\pm$       1.53  &         0.13   \\
 MSWOPs  & -1.03$\pm$   0.79  &    0.08  &          2.31  $\pm$       1.32  &         0.19   \\
\hline
\end{tabular}
\end{scriptsize}
\end{table}
  
 Left panel in Figure~\ref{tc_plots_giants} shows  that when all elements are considered
 there seems to be no difference in the chemical behaviour of the planet host samples
 with respect to their respective comparison samples. This result holds independently of the 
 evolutionary state of the stars (giant, subgiant or main-sequence), showing stars with
 and without planets very similar slopes.
 We note however that the slopes of MS and subgiant stars tend, within the errors,
 to be consistent with zero, with moderate $p$-values.
 Giants, on the other hand, show clearly negative slopes and statistically
 significant low $p$-values.

 The [X/Fe] vs. T$_{\rm C}$ trends when only refractory elements are considered
 are shown in the right panel of Figure~\ref{tc_plots_giants}.
 We note a slight change in 
 behaviour between MSWPs (negative slope) with respect to MSWOPs (positive slope).
 This tendency is also seen in the samples of subgiant stars
 (i.e., negative slopes in planet hosts, positive slopes in non-planet hosts).
 The statistical significance of these trends is however low, with moderate
 $p$-values. 
 When considering giant stars both GWP and GWOP samples show 
 negative [X/Fe] vs. T$_{\rm C}$ trends,  being the slope
 of the GWOP sample slightly more negative. 
 Note that the slopes obtained for the giant samples are statistical significant
 but that is not the case for the MS and subgiant samples.

  We also note that at this point we are considering the sample of giants as a whole,
 i.e., without any mass differentiation, despite the reported difference
 in the metallicity behaviour between stars with masses lower and
 larger than 1.5 M$_{\odot}$. We will analyse the mass segregation
 in detail in Section~\ref{mass_segregation}. 
 We caution that other effects (e.g. uncertainties in the stellar mass determination,
 or the criteria used to discern subgiants from giants)
 may also be present.

\subsection{Trends with evolutionary properties}

\begin{table*}
\centering
\caption{Results from the Spearman's correlation test between the  T$_{\rm C}$ slopes
 and different stellar properties. MC stands for Monte Carlo simulation, CA for
 ``classical'' analysis, SR for Spearman's correlation rank coefficient, ZS means z-score, and
 $p$ denotes the significance of the SR coefficient.}
\label{master_correlation_table}
\begin{tabular}{ll|cccc|cccc}
\hline\noalign{\smallskip}
          &        & \multicolumn{4}{c}{\bf All elements}                   & \multicolumn{4}{c}{\bf Only refractory}                \\
          &        & \multicolumn{2}{c}{\it MC} &\multicolumn{2}{c}{\it CA} & \multicolumn{2}{c}{\it MC} &\multicolumn{2}{c}{\it CA} \\  
Parameter & Sample      &     SR               & ZS     &  SR       & $p$          & SR  & ZS &  SR & $p$                               \\
\hline
[Fe/H]	&	All	&	0.29	$\pm$	0.05	&	2.29	$\pm$	0.45	&	0.31	&	$\sim$ 10$^{\rm -8}$	&	-0.32	$\pm$	0.05	&	-2.57	$\pm$	0.45	&	-0.37	&	$\sim$ 10$^{\rm -12}$	\\
	&	Giants	&	0.44	$\pm$	0.09	&	2.10	$\pm$	0.47	&	0.48	&	$\sim$ 10$^{\rm -7}$	&	-0.42	$\pm$	0.09	&	-1.96	$\pm$	0.45	&	-0.51	&	$\sim$ 10$^{\rm -8}$	\\
	&	Subgiants	&	0.49	$\pm$	0.15	&	1.30	$\pm$	0.49	&	0.52	&	$\sim$ 10$^{\rm -3}$	&	-0.55	$\pm$	0.15	&	-1.48	$\pm$	0.52	&	-0.60	&	$\sim$ 10$^{\rm -4}$	\\
	&	MS	&	0.15	$\pm$	0.07	&	0.92	$\pm$	0.43	&	0.16	&	0.02	&	-0.28	$\pm$	0.07	&	-1.69	$\pm$	0.45	&	-0.32	&	$\sim$ 10$^{\rm -6}$	\\
\hline																											
$\log$g	&	All	&	-0.16	$\pm$	0.05	&	-1.27	$\pm$	0.44	&	-0.18	&	$\sim$ 10$^{\rm -3}$	&	0.22	$\pm$	0.05	&	1.75	$\pm$	0.44	&	0.28	&	$\sim$ 10$^{\rm -7}$	\\
	&	Giants	&	0.26	$\pm$	0.10	&	1.17	$\pm$	0.45	&	0.28	&	$\sim$ 10$^{\rm -3}$	&	0.17	$\pm$	0.10	&	0.74	$\pm$	0.43	&	0.22	&	0.02	\\
	&	Subgiants	&	0.30	$\pm$	0.17	&	0.75	$\pm$	0.45	&	0.33	&	0.06	&	-0.33	$\pm$	0.17	&	-0.82	$\pm$	0.45	&	-0.36	&	0.04	\\
	&	MS	&	-0.17	$\pm$	0.07	&	-1.00	$\pm$	0.41	&	-0.19	&	$\sim$ 0.01	&	0.16	$\pm$	0.07	&	0.98	$\pm$	0.44	&	0.21	&	$\sim$ 10$^{\rm -3}$	\\
\hline																											
M$_{\star}$	&	All	&	0.34	$\pm$	0.05	&	2.70	$\pm$	0.43	&	0.36	&	$\sim$ 10$^{\rm -11}$	&	-0.37	$\pm$	0.05	&	-2.97	$\pm$	0.45	&	-0.45	&	$\sim$ 10$^{\rm -17}$	\\
	&	Giants	&	0.43	$\pm$	0.08	&	2.00	$\pm$	0.43	&	0.50	&	$\sim$ 10$^{\rm -8}$	&	-0.57	$\pm$	0.07	&	-2.84	$\pm$	0.48	&	-0.70	&	$\sim$ 10$^{\rm -17}$	\\
	&	Subgiants	&	0.29	$\pm$	0.17	&	0.71	$\pm$	0.42	&	0.32	&	0.08	&	-0.23	$\pm$	0.19	&	-0.55	$\pm$	0.47	&	-0.25	&	0.17	\\
	&	MS	&	0.31	$\pm$	0.07	&	1.87	$\pm$	0.45	&	0.33	&	$\sim$ 10$^{\rm -6}$	&	-0.32	$\pm$	0.07	&	-1.96	$\pm$	0.46	&	-0.39	&	$\sim$ 10$^{\rm -8}$	\\
\hline																											
Age	&	All	&	-0.11	$\pm$	0.06	&	-0.82	$\pm$	0.43	&	-0.14	&	$\sim$ 0.01	&	0.25	$\pm$	0.05	&	1.94	$\pm$	0.44	&	0.31	&	$\sim$ 10$^{\rm -8}$	\\
	&	Giants	&	-0.35	$\pm$	0.09	&	-1.62	$\pm$	0.43	&	-0.47	&	$\sim$ 10$^{\rm -7}$	&	0.52	$\pm$	0.08	&	2.52	$\pm$	0.49	&	0.65	&	$\sim$ 10$^{\rm -14}$	\\
	&	Subgiants	&	-0.31	$\pm$	0.16	&	-0.76	$\pm$	0.42	&	-0.35	&	0.05	&	0.10	$\pm$	0.20	&	0.23	$\pm$	0.49	&	0.11	&	0.54	\\
	&	MS	&	-0.02	$\pm$	0.07	&	-0.10	$\pm$	0.42	&	-0.03	&	0.72	&	0.15	$\pm$	0.07	&	0.86	$\pm$	0.42	&	0.17	&	0.02	\\
\hline																											
R$_{\star}$	&	All	&	0.23	$\pm$	0.05	&	1.79	$\pm$	0.44	&	0.25	&	$\sim$ 10$^{\rm -6}$	&	-0.27	$\pm$	0.05	&	-2.09	$\pm$	0.45	&	-0.33	&	$\sim$ 10$^{\rm -9}$	\\
	&	Giants	&	-0.11	$\pm$	0.10	&	-0.48	$\pm$	0.45	&	-0.11	&	0.27	&	-0.31	$\pm$	0.09	&	-1.41	$\pm$	0.43	&	-0.39	&	$\sim$ 10$^{\rm -5}$	\\
	&	Subgiants	&	-0.19	$\pm$	0.19	&	-0.44	$\pm$	0.46	&	-0.22	&	0.22	&	0.13	$\pm$	0.18	&	0.32	$\pm$	0.43	&	0.14	&	0.43	\\
	&	MS	&	0.29	$\pm$	0.07	&	1.74	$\pm$	0.45	&	0.31	&	$\sim$ 10$^{\rm -5}$	&	-0.24	$\pm$	0.07	&	-1.41	$\pm$	0.46	&	-0.29	&	$\sim$ 10$^{\rm -5}$	\\
\hline																											
 d$_{\rm galact}$	&	All	&	0.00	$\pm$	0.08	&	0.03	$\pm$	0.44	&	0.00	&	0.99	&	-0.21	$\pm$	0.07	&	-1.20	$\pm$	0.44	&	-0.22	&	$\sim$ 10$^{\rm -3}$	\\
\hline
\end{tabular}
\end{table*}

  In the previous subsection we have found that a different chemical trend
  may exist between planets and non-planets hosts when only refractory
  elements are considered, but that this difference seem to be only
  present in MS and subgiant stars, and not in giants.
  This behaviour resembles the gas-giant planet-metallicity correlation
  known to hold for MS and subgiant stars but controversial
  when considered giants (e.g. MA13).
  It is, therefore, natural to ask whether the
  obtained  abundance trends  correlate
  with other evolutionary parameter.

  We have thus performed a search for correlations between the derived
  T$_{\rm C}$-slopes for each individual star and the evolutionary
  parameters namely- surface gravity, stellar mass, age, and radius.
  Stellar metallicity and the stellar mean galactocentric distance (d$_{\rm galact}$) have been as well considered
  with values taken from our own previous work and from \cite{2011A&A...530A.138C} respectively. 
  Two kinds of analysis have been performed. The first one consists on the
  classical Spearman's correlation test. 
  Further analysis includes the evaluation of the significance
  of the correlations by a bootstrap Monte Carlo (MC) test plus a Gaussian random shift
  of each data-point within its error bars. The tests were done using the code
  {\sc MCSpearman}\footnote{https://github.com/PACurran/MCSpearman/} by
  \citet{2014arXiv1411.3816C} and the results are shown in Table~\ref{master_correlation_table}.
  It is clear from this table that the classical
  analysis suggests moderate but highly significant correlations between
  the T$_{\rm C}$-slopes and the evolutionary parameters. 
  The MC simulations do not exclude such dependencies, however suggest
  that the correlations are weak, being the z-score values in all cases lower
  than 3$\sigma$.

  Our analysis show that the derived T$_{\rm C}$-slope correlates
  with the stellar metallicity. Such a correlation suggests that
  Galactic Chemical Evolution  (GCE) effects may be impacting our derived
  abundance patterns. While some authors \citep{2013A&A...552A...6G,2014A&A...564L..15A}
  have tried to account for these effects by fitting straight lines to the [X/Fe]
  vs. [Fe/H] plots, others \citep[e.g.][]{2014A&A...561A...7R} argue that correcting from GCE effects in this manner
  may prevent us from finding elemental depletions due to planet formation. 
  
\begin{figure*}[!htb]
\centering
\begin{minipage}{0.30\linewidth}
\includegraphics[angle=270,scale=0.35]{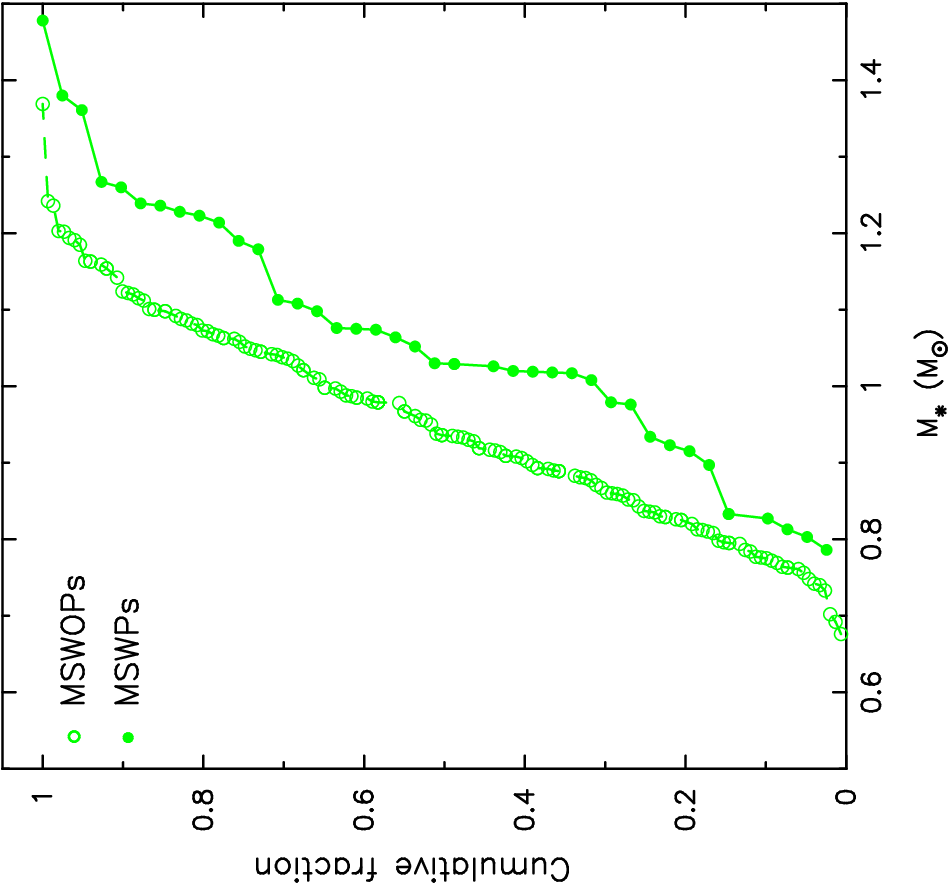}
\end{minipage}
\begin{minipage}{0.30\linewidth}
\includegraphics[angle=270,scale=0.35]{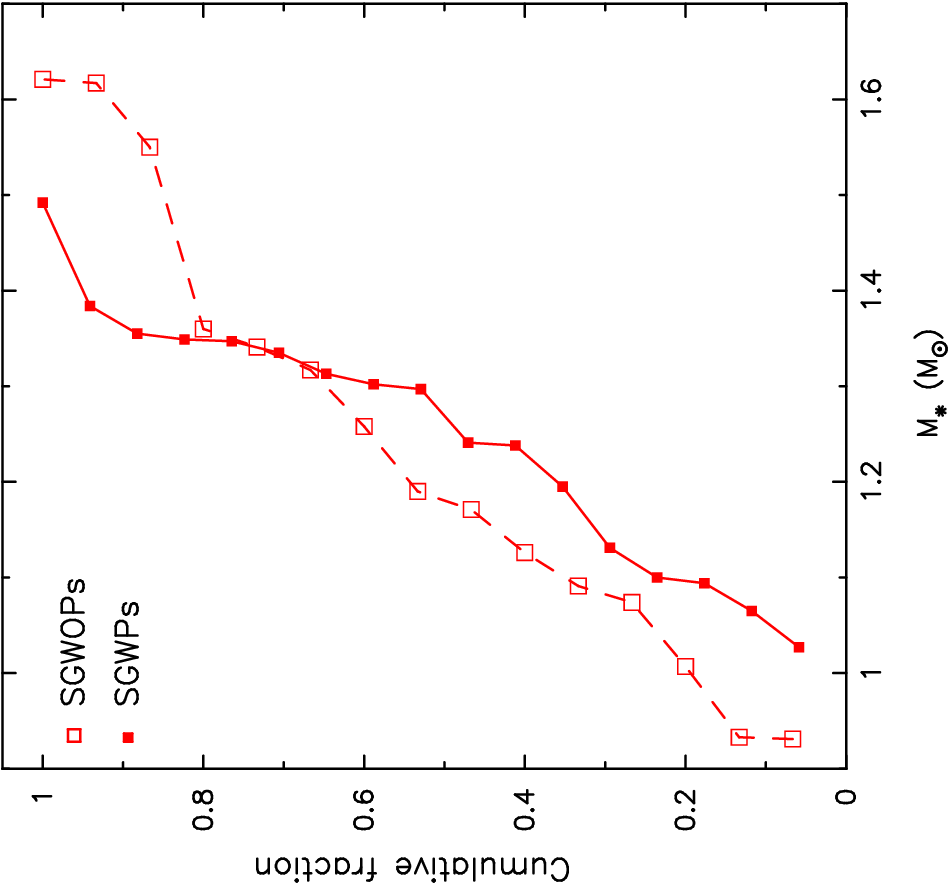}
\end{minipage}
\begin{minipage}{0.30\linewidth}
\includegraphics[angle=270,scale=0.35]{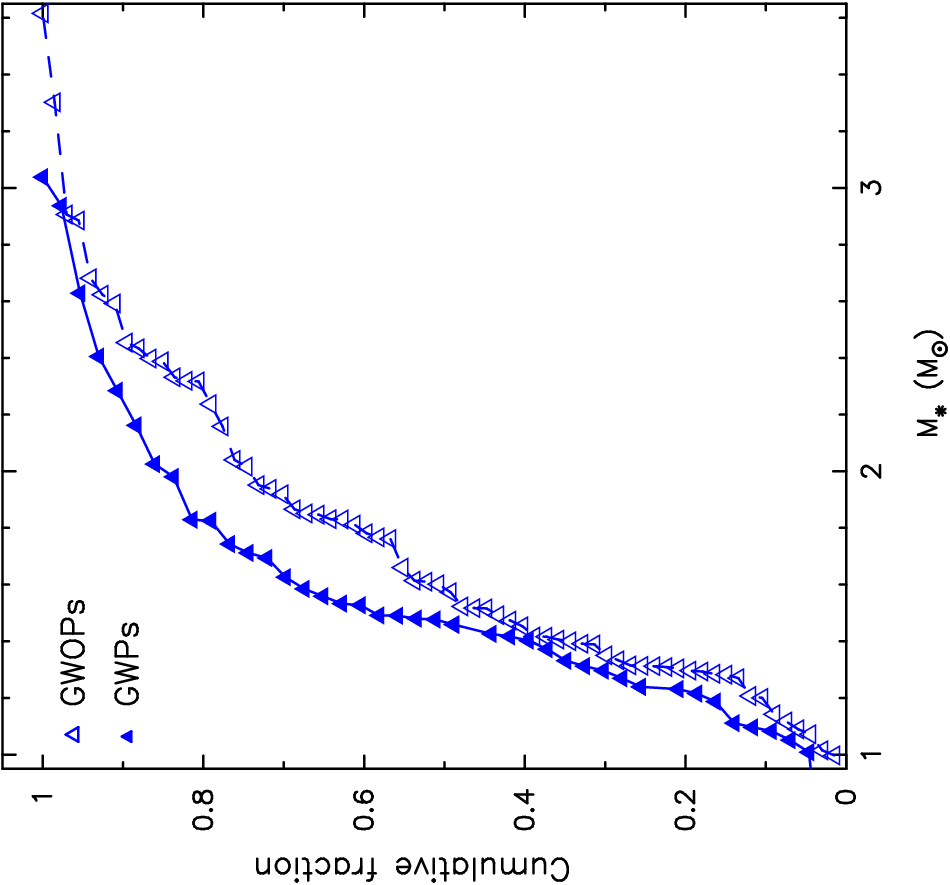}
\end{minipage}
\caption{
 Cumulative distribution function of stellar masses for MS stars (left),
 subgiants (middle), and giants (right).}
\label{distribuciones_masas}
\end{figure*}

\begin{figure*}[!htb]
\centering
\begin{minipage}{0.30\linewidth}
\includegraphics[angle=270,scale=0.35]{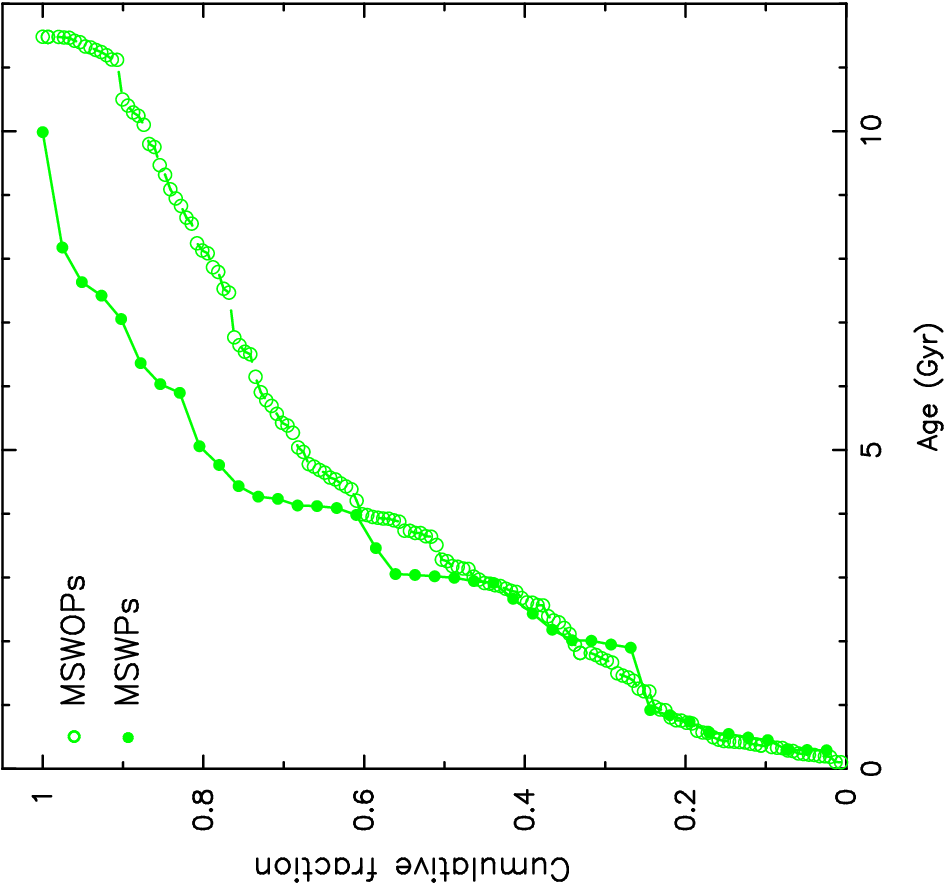}
\end{minipage}
\begin{minipage}{0.30\linewidth}
\includegraphics[angle=270,scale=0.35]{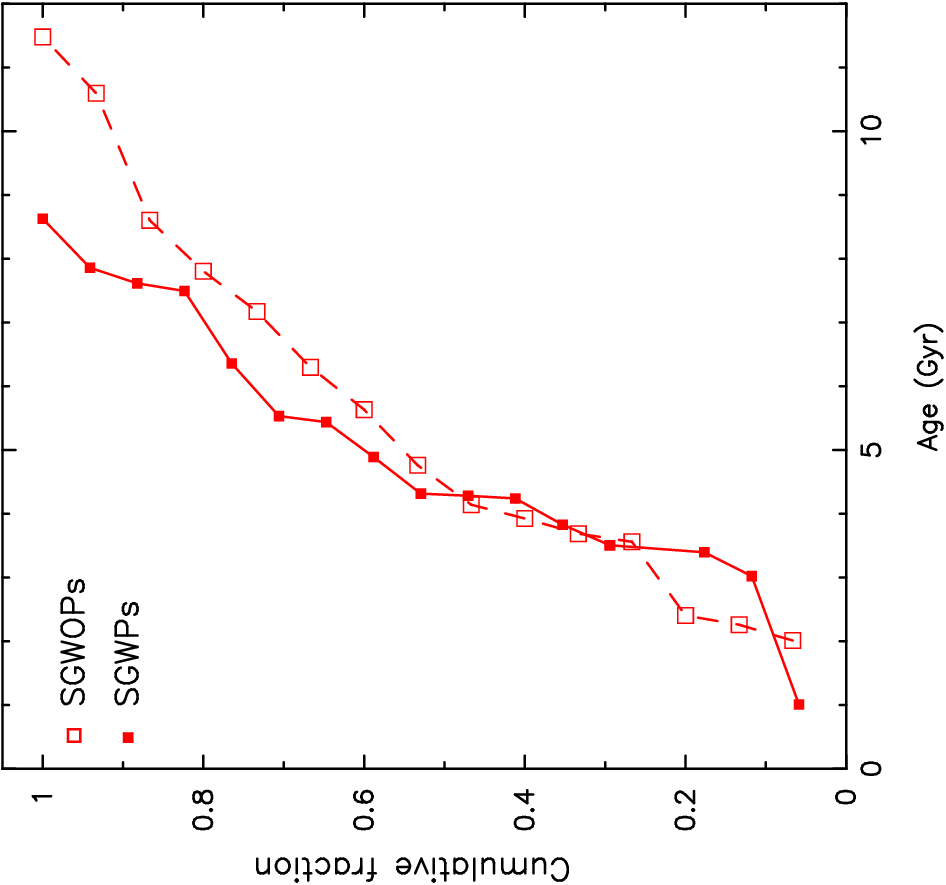}
\end{minipage}
\begin{minipage}{0.30\linewidth}
\includegraphics[angle=270,scale=0.35]{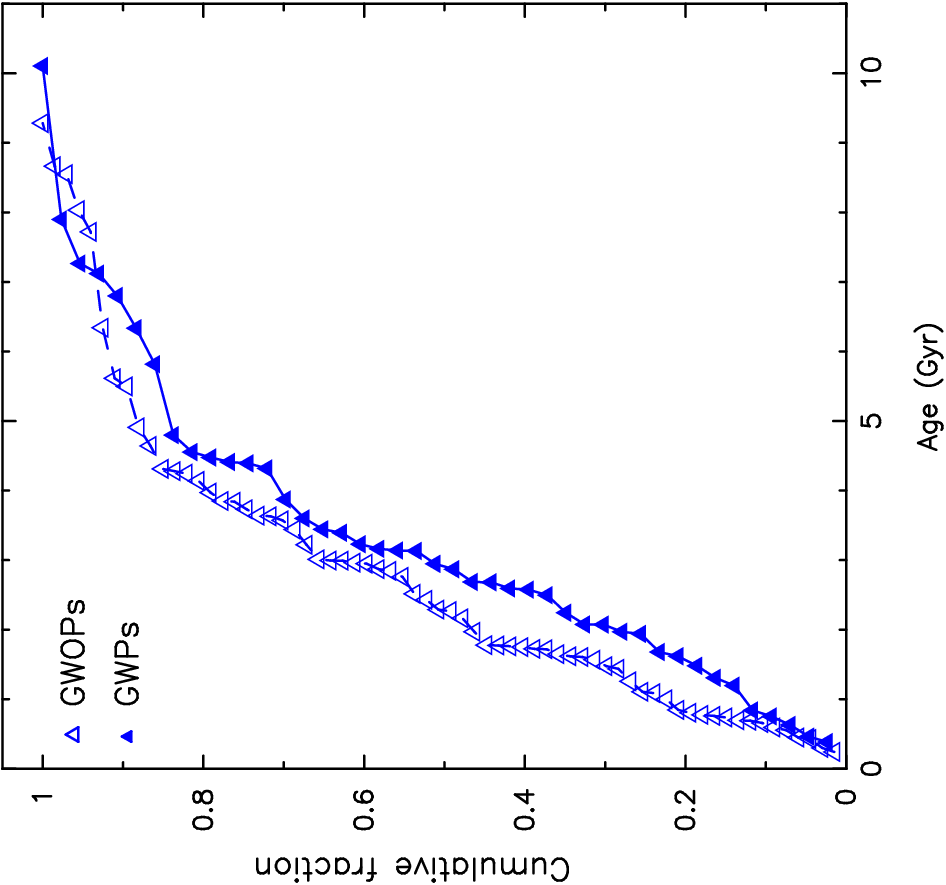}
\end{minipage}
\caption{
 Cumulative distribution function of the stellar age for MS stars (left),
 subgiants (middle), and giants (right).}
\label{distribuciones_ages}
\end{figure*}

  Table~\ref{master_correlation_table} also shows a clear correlation between the 
  T$_{\rm C}$-slope and the stellar mass, or the stellar age (see the corresponding
  $p$-values).  We note that the correlations were performed using all stars
  (planet and non planet hosts) together. 
  Less massive and older stars show more positive T$_{\rm C}^{\rm ref}$-slopes,
  and more negative T$_{\rm C}^{\rm all}$-slopes.
  This result agree with recent studies of solar twins in which the existence
  of a correlation between [X/Fe] and the stellar age have been 
  revealed \citep{2015A&A...579A..52N,2015arXiv151101012S}.
  Following this line of reasoning, a comparison of the stellar masses and ages
  between planet hosts and non planet hosts were performed.
  Figure~\ref{distribuciones_masas} 
  and ~\ref{distribuciones_ages}
  show the corresponding cumulative distribution functions, while some statistical
  diagnostics are presented in Table~\ref{mass_age_table}.
  These figures show that there seems to be a hint of MS and subgiant non-planet hosts
  to have slightly smaller masses and older ages than planet hosts.
  This could be a selection effect as radial velocity surveys tend to 
  target stars with low levels of activity. 
  We note that for the case of giants the behaviour seems to be the opposite,
  being GWOPs slightly younger and massive than GWPs.

  In order to test the statistical significance of these trends
  several KS tests were performed (Table~\ref{ksteststable}).
  The results from the KS test show that the differences
  in mass or age between planet and non-planet hosts
  are in general not significant from an statistical
  point of view. 
  The mass segregation between
  planets and non-planet hosts in MS stars appears to be the only trend
  that might be statistically significant. 
  
   In order to check whether GCE effects my affect or not our results, abundances
  were corrected by fitting straight lines to the [X/Fe] vs. [Fe/H] plots
  (see Figure~\ref{swds_swods_xfe_vs_feh}).
  As before,  T$_{\rm C}$-slopes were computed for each individual star and
  a search for correlations was performed. We find that most of the
  correlations with the evolutionary parameters remain.

\begin{table}
\centering
\caption{Stellar mass and stellar age statistics of the stellar samples.}
\label{mass_age_table}
\begin{tabular}{lccccc}
\hline\noalign{\smallskip}
\multicolumn{6}{c}{Stellar mass (M$_{\odot}$)}\\
\hline
Sample   &      Mean     &      Median   &      $\sigma$         &      Min.     &      Max. \\
\hline
MSWPs    &      1.06     &      1.03     &      0.16     &      0.79     &      1.48 \\
MSWOPs   &      0.95     &      0.94     &      0.14     &      0.68     &      1.37 \\
SGWPS    &      1.25     &      1.30     &      0.13     &      1.03     &      1.49 \\
SGWOPs   &      1.24     &      1.19     &      0.22     &      0.93     &      1.62 \\
GWPs     &      1.60     &      1.48     &      0.48     &      1.01     &      3.04 \\
GWOPs    &      1.76     &      1.60     &      0.56     &      1.00     &      3.62 \\
\hline
\multicolumn{6}{c}{Stellar age (Gyr)}\\
\hline
Sample   &      Mean     &      Median   &      $\sigma$         &      Min.     &      Max. \\
\hline
MSWPs    &      3.41     &      3.02     &      2.41     &      0.29     &      9.99 \\
MSWOPs   &      4.32     &      3.28     &      3.62     &      0.10     &      11.48 \\
SGWPs    &      5.00     &      4.32     &      1.98     &      1.01     &      8.63 \\
SGWOPs   &      5.62     &      4.76     &      2.88     &      2.01     &      11.48 \\
GWPs     &      3.37     &      2.95     &      2.15     &      0.38     &      10.10 \\
GWOPs    &      2.82     &      2.29     &      2.15     &      0.24     &      9.28 \\
\hline
\end{tabular}
\end{table}

\begin{table}
\centering
\caption{Results of the K-S tests performed in this work. We consider a
confidence level of 98\% in order to reject the null hypothesis H$_{0}$
(both samples coming from the same underlying continuous
distribution).}
\label{ksteststable}
\begin{scriptsize}
\begin{tabular}{lcccccc}
\hline\noalign{\smallskip}
\multicolumn{7}{c}{Stellar mass}\\
\hline
Sample  &  n$_{\rm planets}$ & n$_{\rm comparison}$ & n$_{\rm eff}$ & $D$  & $p$   & H$_{0}^{\ddag}$ \\
\hline
MS	 &	41	&	151	&	32	&	0.36	&	$\sim$10$^{\rm -4}$ & 1 \\
Subgiants&	17	&	15	&	7	&	0.24	&	0.69		    & 0 \\	
Giants   &	43	&	67	&	26	&	0.22	&	0.15		    & 0 \\ 	
\hline
\multicolumn{7}{c}{Stellar age}\\
\hline
Sample  &  n$_{\rm planets}$ & n$_{\rm comparison}$ & n$_{\rm eff}$ & $D$  & $p$   & H$_{0}^{\ddag}$ \\
\hline
MS	&	41      &       151     &       32      & 0.18 & 0.24 & 0 \\
Subgiants&      17      &       15      &       7       & 0.17 & 0.95 & 0 \\
Giants&          43      &       67      &       26      & 0.22 & 0.15 & 0 \\
\hline
\end{tabular}
\tablefoot{ 
 $D$ is the maximum deviation between the empirical distribution
 function of samples 1 and 2. $p$ corresponds to the estimated
 likelihood of the null hypothesis, a  value that is known to be reasonably
 accurate for sample sizes for which n$_{\rm eff}$ $\ge$ 4.
 $^{\ddag}$ (0): Accept null hypothesis; (1): Reject null hypothesis.}
\end{scriptsize}
\end{table}

  Finally, we have redone our analysis but considering only
  those stars similar to our Sun (the so-called solar analogs). Results are given
  in Table~\ref{master_corr_solar}. We note that the correlations discussed
  before are not so evident.
  However, the classical analysis suggests
  that some correlations may still be present. In particular between
  T$^{\rm all}_{\rm C}$ and $\log$g, and perhaps the stellar age,
  and between T$^{\rm ref}_{\rm C}$ and [Fe/H].
  A more detailed strictly differential analysis of these stars
  should be done to clarify this point and to properly
  compare with previous results \citep[e.g.][]{2014A&A...564L..15A,2015A&A...580A..24D}.


\begin{table*}
\centering
\caption{  Same as Table~\ref{master_correlation_table} but for a sample of
 34 solar analogs (T$_{\rm eff}$ = 5777 $\pm$ 200 K, $\log$g = 4.44 $\pm$  0.20
 dex, [Fe/H] = 0.00 $\pm$  0.20 dex).} 
\label{master_corr_solar}
\begin{tabular}{l|cccc|cccc}
\hline\noalign{\smallskip}
          & \multicolumn{4}{c}{\bf All elements}                   & \multicolumn{4}{c}{\bf Only refractory}                \\
          & \multicolumn{2}{c}{\it MC} &\multicolumn{2}{c}{\it CA} & \multicolumn{2}{c}{\it MC} &\multicolumn{2}{c}{\it CA} \\
Parameter &     SR               & ZS     &  SR       & $p$          & SR  & ZS &  SR & $p$                               \\
\hline
[Fe/H]	&	0.07	$\pm$	0.18	&	0.18	$\pm$	0.44	&	0.06	&	0.72	&	-0.38	$\pm$	0.15	&	-0.96	$\pm$	0.43	&	-0.48	&	$\sim$ 10$^{\rm -3}$	\\
$\log$g	&	0.32	$\pm$	0.15	&	0.79	$\pm$	0.40	&	0.44	&	0.01	&	-0.02	$\pm$	0.17	&	-0.05	$\pm$	0.42	&	0.04	&	0.81	\\
M$_{\star}$	&	0.13	$\pm$	0.17	&	0.32	$\pm$	0.42	&	0.13	&	0.47	&	-0.18	$\pm$	0.18	&	-0.45	$\pm$	0.44	&	-0.22	&	0.20	\\
Age	&	-0.22	$\pm$	0.16	&	-0.53	$\pm$	0.40	&	-0.29	&	0.10	&	0.08	$\pm$	0.18	&	0.20	$\pm$	0.43	&	0.06	&	0.72	\\
R$_{\star}$	&	-0.16	$\pm$	0.18	&	-0.38	$\pm$	0.43	&	-0.20	&	0.26	&	0.11	$\pm$	0.18	&	0.26	$\pm$	0.43	&	0.09	&	0.62	\\
 d$_{\rm galact}$	&	-0.04	$\pm$	0.18	&	-0.10	$\pm$	0.43	&	-0.05	&	0.77	&	-0.13	$\pm$	0.18	&	-0.33	$\pm$	0.46	&	-0.15	&	0.39	\\
\hline
\end{tabular}
\end{table*}

\subsection{Mass segregation in giants and abundance trends}
\label{mass_segregation}

 MA13 find that for giant stars as a whole there is no correlation between the presence of giant planets and the metallicity of
 the star. However, within the lack of correlation there seems to be hidden a dependency on the stellar mass.
 While the less massive giant stars with planets
 (M$_{\star}$ $\le$ 1.5 M$_{\odot}$) are not metal rich, 
 the metallicity of the sample of massive (M$_{\star}$ > 1.5 M$_{\odot}$ ) giant stars with
 planets is higher than that of a similar sample of stars without planets.
 It is therefore natural to ask whether there are no differences in the
 <[X/Fe]>-T$_{\rm C}$ trends between stars more massive than 1.5 M$_{\odot}$ and 
 less massive giants. 

\begin{figure*}
\centering
\begin{minipage}{0.49\linewidth}
\includegraphics[angle=270,scale=0.375]{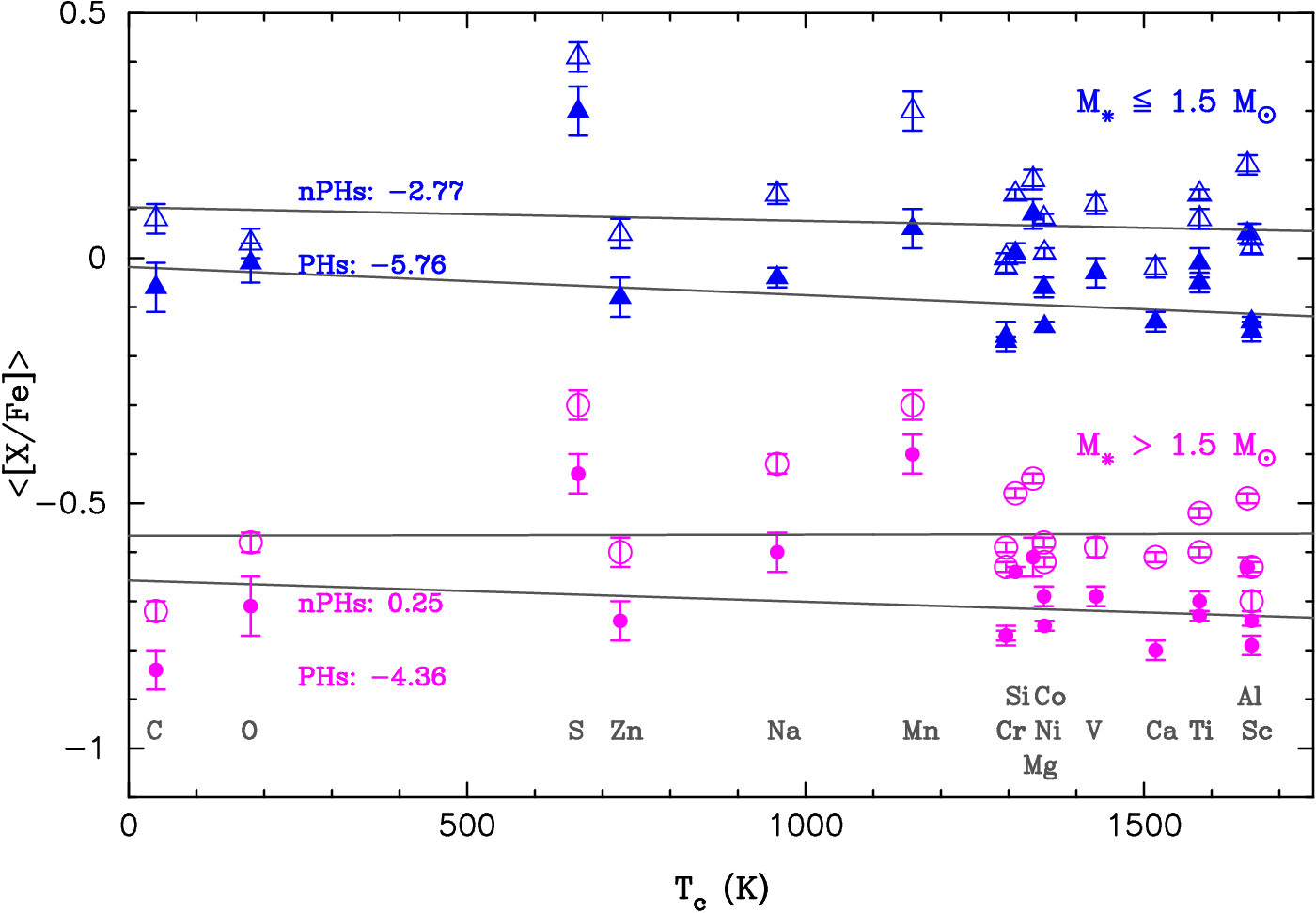}
\end{minipage}
\begin{minipage}{0.49\linewidth}
\includegraphics[angle=270,scale=0.375]{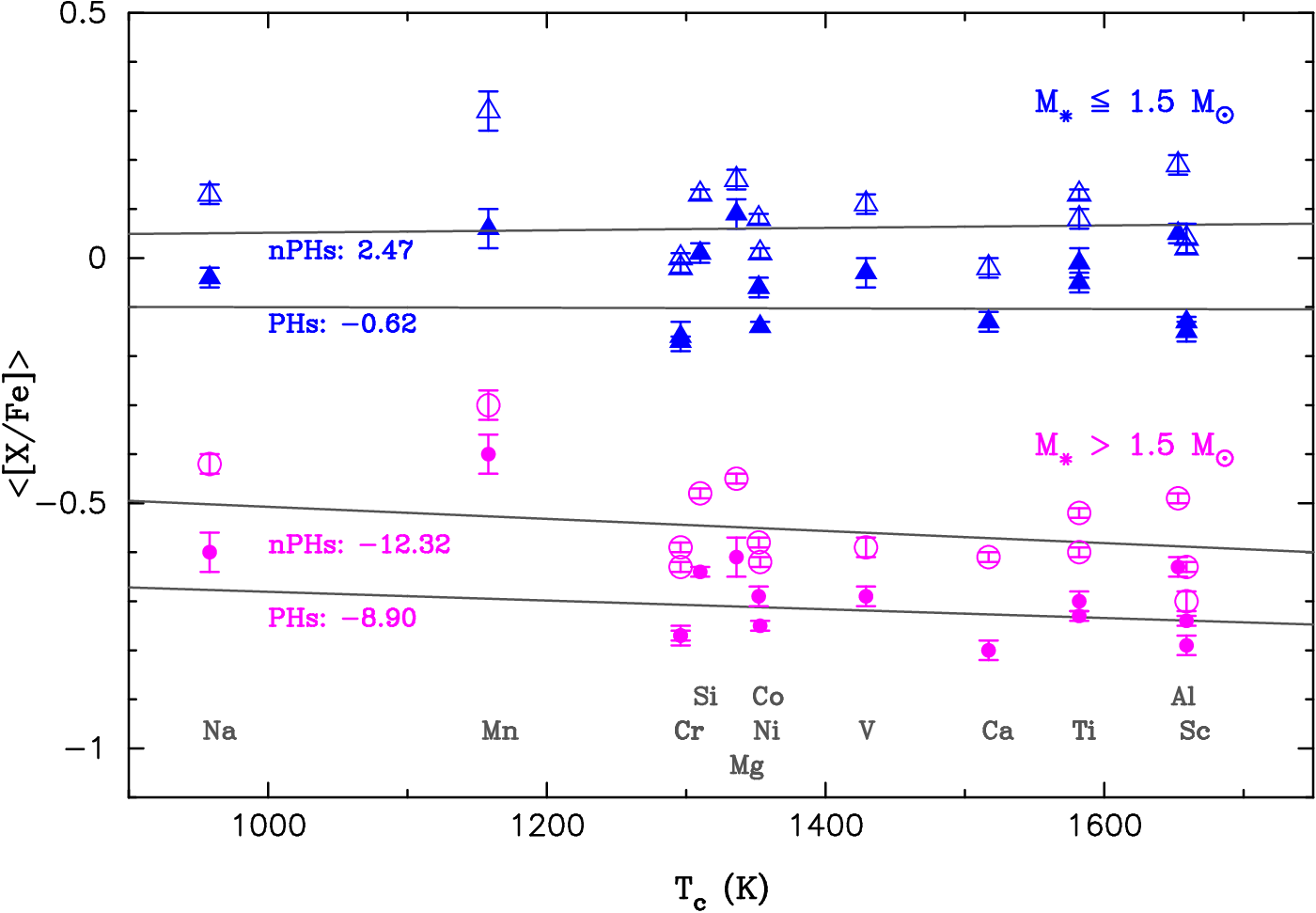}
\end{minipage}
\caption{
  <[X/Fe]>-T$_{\rm C}$ trends for the giant stars.
  GWPs with M$_{\star}$ $>$ 1.5 M$_{\odot}$
  are plotted with purple circles, GWPs less massive than  1.5 M$_{\odot}$ in blue triangles.
  Filled symbols indicate planet hosts.
  Each planet host subsample is shown against its corresponding comparison subsample
  with an offset of -0.15 for the sake of clarity.
  The offset between the sample of giants with M$_{\star}$ $>$ 1.5 M$_{\odot}$ and 
  less massive giants is -0.75.
  For guidance, the derived slopes are shown
  in the plots (units of 10$^{\rm -5}$ dex/K).
  The left panel shows the   <[X/Fe]>-T$_{\rm C}$ trends
  when all elements
  (volatiles and refractories) are taken into account whilst the
  right one shows the <[X/Fe]>-T$_{\rm C}$ trends when only refractories are considered.
}
\label{tc_plots_giants_masas}
\end{figure*}

 Figure~\ref{tc_plots_giants_masas} shows the mean <[X/Fe]>-T$_{\rm C}$ trend of
 GWPs and GWOPs for giants with M$_{\star}$ $\le$ 1.5 M$_{\odot}$ and
 giants in the mass domain M$_{\star}$ > 1.5 M$_{\odot}$. The results of the
 corresponding linear fits are given in Table~\ref{linearfits_giants}.
 We find that when considering all elements, the slopes are always negative
 with the only exception of the GWOP sample for M$_{\star}$ $>$ 1.5 M$_{\odot}$.
 However, we note that in this case the slope is consistent with zero.
 When considering only refractory elements, for the more massive giants,
 stars with and without planets show similar negative slopes.
 For giants in the mass domain M$_{\star}$ $\le$ 1.5 M$_{\odot}$,
 GWPs show a slightly negative slope, whilst GWOP a slightly
 positive one. However, we note that both slopes within their
 corresponding errors are compatible with zero.
 We conclude that giant stars do not show differences in the  <[X/Fe]>-T$_{\rm C}$ trends
 between planet and not planet hosts. This result holds independently of whether
 all giants are considered (Section~\ref{xfe_tc_trends})
 or we separate the sample according to the stellar mass.

\begin{table}
\centering
\caption{Results of the  <[X/Fe]>-T$_{\rm C}$ linear fits for the giant stars
 according to their masses. For each fit its
 probability of slope ``being by chance'' ($p$) is also given.}
\label{linearfits_giants}
\begin{scriptsize}
\begin{tabular}{lcccc}
\hline\noalign{\smallskip}
Sample   & \multicolumn{2}{c}{\bf All elements}               & \multicolumn{2}{c}{\bf Only refractory}        \\
         & slope ($\times$10$^{\rm -5}$dex/K)       & $p$.    & slope ($\times$10$^{\rm -5}$dex/K)  &  $p$     \\
\hline
 \multicolumn{5}{c}{M$_{\star}$ $>$ 1.5 M$_{\odot}$} \\
\hline
 GWPs   & -4.36  $\pm$ 1.58   & 0.08    &   -8.90        $\pm$  2.42       &   0.06         \\
 GWOPs  & 0.25   $\pm$ 0.89   & 0.05    &   -12.32       $\pm$  1.82       &   0.05         \\
\hline
 \multicolumn{5}{c}{M$_{\star}$ $\le$ 1.5 M$_{\odot}$} \\
\hline
 GWPs   & -5.76 $\pm$ 1.58     & 0.09    &  -0.62         $\pm$ 2.35       &  0.10          \\
 GWOPs  & -2.77 $\pm$ 1.20     & 0.08    &   2.47         $\pm$ 2.12       &  0.07          \\
\hline
\end{tabular}
\end{scriptsize}
\end{table}

 We note that giants in the mass domain M$_{\star}$ $\le$ 1.5 M$_{\odot}$
 show more positive <[X/Fe]>-T$_{\rm C}^{ref}$ slopes than giants
 with M$_{\star}$ > 1.5 M$_{\odot}$. This fact could in principle
 be explained by the anticorrelation between the stellar mass
 and the T$_{\rm C}^{ref}$-slope seen in the previous subsection.
 Further,
 could be hidden within as MA13 pointed out that giants in the
 mass domain M$_{\star}$ $\le$ 1.5 M$_{\odot}$
 show lower metallicities than giants with  M$_{\star}$ > 1.5 M$_{\odot}$.
 Note that the T$_{\rm C}^{ref}$-slope was shown before to depend on the stellar metallicity.


\section{Discussion}
\label{seccion_discussion}

  In the previous section we have shown that stellar T$_{\rm C}$-slopes correlate with 
  the stellar evolutionary parameters.
  The data suggest that there might be a different behaviour in the <[Fe/H]>-T$_{\rm C}$ trends
  between planets and non planet hosts for MS, and subgiant stars. 
  However, there seem to be no difference between planet and non planet hosts among
  the sample of giants.
  
  The finding that the MS non-planet hosts of our sample are less massive and  perhaps older
  than the planet hosts is,  if significant, somehow surprising, and might hide some bias
  in the subsample selection. 
  In fact, spectroscopic targets for planet searches are often deliberately chosen in order
  to be slow rotators and typically inactive which should sample a population of MS stars older
  than the average population of the same spectral type.
  Recently, \cite{2015A&A...575A..18B} have analysed whether exoplanet hosts are peculiar
  with respect to field stars not hosting planets regarding ages and found that both
  samples are homogeneous within the solar neighbourhood with a median age distribution of
  4.8 Gyr slightly older than the average thin disc population. 
  This seems to be at odds with our results, although it may be an effect of the
  sample selection. Note that our MS stars are younger on average
  (See Table~\ref{mass_age_table}). Further,  the only difference
  between MSWPs and MSWOPs found to be statistically significant
  is in the stellar mass.

  \cite{2009ApJ...698L...1H} suggested that the observed correlation
  between the presence of gas-giant planets and enhanced stellar metallicity
  observed in MS planet hosts might be related to a possible inner-disc
  origin of these stars. 
  The fact that stars with low-mass planets do not show the metal-rich signature
  does not necessary contradict this idea, although further investigations
  are needed to clarify this point.
  Radial mixing is a secular process, and its effect is known to increase with time,
  older stars migrate further and come from a region with significantly
  different abundances. On the other hand, age and mass are related quantities.
  MS non-planet host may show a different [X/Fe]-T$_{\rm C}$ trends
  (with respect to planet host) just simply
  because these stars are slightly older and less massive, and possibly more
  contaminated by stars from the outer disc. 
  Stars from the outer disc  (at larger galactocentric distances) 
  are expected to show lower metallicities \citep[e.g.][Figure~5]{2008A&A...490..613L}
  and therefore  larger [X/Fe] values for most elements (see Figure~\ref{swds_swods_xfe_vs_feh}),
  which may explain their positive T$^{\rm ref}_{\rm C}$ slopes. 
  In this framework, the lack of a difference between GWPs and GWOPs is explained by the 
  fact that these samples are younger and more massive than their MS counterparts,
  and therefore significantly less affected by radial mixing.

  Along this line, hints of a correlation between the T$_{\rm C}$ slopes
  and the stellar age have already been reported
  \citep{2014A&A...564L..15A,2015A&A...579A..20M,2015A&A...579A..52N,2015arXiv151101012S}.
  \cite{2014A&A...564L..15A} use the stellar mean galactocentric distance
  as a proxy of the stellar birthplace finding tentative evidence of 
  a correlation with the T$_{\rm C}$ slope. Such a correlation
  seem to be also present in our data (although when considering
  only refractory elements, see Table~\ref{master_correlation_table}).
  Unfortunately, no values of d$_{\rm galact}$ are available for most of our giant stars.


  An alternative  interpretation of the  <[Fe/H]>-T$_{\rm C}$ abundance patterns
  in planet host was given by \citet[][hereafter ME09]{2009ApJ...704L..66M}.
  ME09 report a deficit of refractory elements in the Sun with respect to other solar twins.
  ME09 
  conclude that the most likely explanation is related to the formation
  of planetary systems like our own, in particular to the formation
  of low-mass rocky planets.

  In the analysis performed in Section~\ref{section_analysis} we 
  have deliberately tried to exclude
  stars with known low-mass planets from the sample. Nevertheless, a difference
  between planet hosts and non-planet host is still present in MS stars in the
  T$_{\rm C}^{\rm ref}$ analysis. This is in line with the results of MA15
  where possible differences in abundance trends were found in stars with
  cool giant-planets, but not in stars with low-mass planets.
  Since the commonly accepted scenario of gas-giant planets formation requires the previous creation
  of a rocky core, the hypothesis that    
  the atmospheres of planet hosts can be contaminated by gas depleted
  in refractories, may still holds. 
  The contamination of gas depleted in refractories due to the planet formation
  process needs very accurate timing as the star needs to retain the 
  protoplanetary disc long enough so the planetary signatures are not cleared out by a deep convection zone on the star.  
  Thus as the star evolves off the MS to become a giant this chemical fingerprint should be erased.
  In principle at the base of the Red Giant Branch phase most of the envelope should be fully convective.
  In this scenario subgiant stars are expected to show similar chemical
  fingerprint of planet formation as MS stars but giant stars should have it erased as planet hosts.

  However, Figure~\ref{tc_plots_giants} (right) shows that the sample
  of stars that  show hints of changing
  its chemical behaviour is the one without planets
  (from positive slopes for MSWOPs and SGWOPs to negative slopes for GWOPs). 
  Thus, other explanations are required to explain this result. The presence
  of a galactic radial mixing
  is in agreement with the fact that we seem to
  be comparing two different populations of stars,
  with the stars in the GWOP sample being more massive and younger
  than stars in the SGWOP and MSWOP samples. 

  It is important to keep in mind that the stars selected for planet searches
  around evolved stars are more massive than their MS counterparts 
   \citep[see e.g.][]{2016A&A...585A..73N} and that it has been shown in MA13
  and \cite{2015A&A...574A.116R} that planet occurrence rate does indeed seem to depend on both, 
 stellar mass and stellar metallicity.
  The different findings for lower and higher mass stars or MS and evolved systems
  does not appear to be simply a consequence of a polluted sample of planet hosts
  with non-planet bearing stars. This explanation put forward
  by \cite{2015A&A...574A.116R}  is what explains their sample of stars but does not
  hold in our analysis of a larger sample (3 times larger) even when we account
  for a possible sample contamination. We find that we are indeed possibly dealing
  with different populations of stars and we hope that improving sample statistics
  in the future will allow to better clean the samples
  to reveal clues on the planet formation process under different conditions.


%
%
%
  
\section{Summary}\label{conclusions}

  In this work a detailed chemical analysis of a large sample of evolved
  (subgiants and red giants) with planets has been presented. Their
  chemical abundances has been compared to those of main-sequence stars.

  No clear difference has been found in <[X/Fe]>-T$_{\rm C}$
  trends between planet and non-planet hosts when all elements are considered
  in the analysis. However, when the analysis is restricted to only
  refractory elements, planet and non-planet hosts might show different
  T$_{\rm C}$-slopes. This result holds for subgiant and giant stars,
  but not for giants.
 
  The data suggest moderate but highly significant correlations between the
  T$_{\rm C}$-slopes and the stellar evolutionary parameters, namely
  stellar mass and age. Less massive and older stars show more
  positive T$^{\rm ref}_{\rm C}$-slopes and more negative T$^{\rm all}_{\rm C}$-slopes.
  In this line, a hint of a difference in terms of
  mass and age seem to be present among our sample of MS stars although
  this result should be further investigated, as it seems to be
  only statistically significant for the stellar mass.
  We had also found that giants with masses M$_{\star}$ $\le$ 1.5 M$_{\odot}$
  show more positive <[X/Fe]>-T$_{\rm C}^{ref}$ slopes than more massive giants,
  in agreement with their lower masses and metallicities.

  Galactic radial mixing offers a suitable scenario for the observed trends.
  Giant stars are more massive and younger than their MS counterparts
   and therefore less contaminated by stars from the outer disc, leading to no chemical
  differences between planet and non-planet hosts.
  On the other hand, less massive and older stars in the MSWOP sample
  may account for different chemical trends between planets and non-planet hosts.
  Other scenarios invoking the formation of planets do not seem to be
  supported by our data.

  Finally, we note that while general trends between the
  T$_{\rm C}$ slopes and evolutionary parameters may be present,
  it does not exclude other processes, such as planetary formation,
  planet engulfment, or dust-gas segregation in protoplanetary
  disc that may affect the stellar  photospheric abundance of refractory elements
  relative to volatiles \citep[][]{2015ApJ...804...40G,2015arXiv151101012S}.


\begin{acknowledgements}
 
  This research was supported by the Italian Ministry of Education,
  University, and Research  through the
  \emph{PREMIALE WOW 2013} research project under grant
  \emph{Ricerca di pianeti intorno a stelle di piccola massa}. 
  E. V. acknowledges support from the \emph{On the rocks} project
  funded by the Spanish Ministerio de Econom\'ia y Competitividad
  under grant \emph{AYA2014-55840-P}.  
  Carlos Eiroa is acknowledged for valuable discussions.

\end{acknowledgements}


\bibliographystyle{aa}
\bibliography{tc_gigantes.bib}

\begin{thebibliography}{71}
\expandafter\ifx\csname natexlab\endcsname\relax\def\natexlab#1{#1}\fi

\bibitem[{{Adibekyan} {et~al.}(2014){Adibekyan}, {Gonz{\'a}lez Hern{\'a}ndez},
  {Delgado Mena}, {Sousa}, {Santos}, {Israelian}, {Figueira}, \& {Bertran de
  Lis}}]{2014A&A...564L..15A}
{Adibekyan}, V.~Z., {Gonz{\'a}lez Hern{\'a}ndez}, J.~I., {Delgado Mena}, E.,
  {et~al.} 2014, \aap, 564, L15

\bibitem[{{Allende Prieto} {et~al.}(2004){Allende Prieto}, {Barklem},
  {Lambert}, \& {Cunha}}]{s4n}
{Allende Prieto}, C., {Barklem}, P.~S., {Lambert}, D.~L., \& {Cunha}, K. 2004,
  \aap, 420, 183

\bibitem[{{Allende Prieto} {et~al.}(2001){Allende Prieto}, {Lambert}, \&
  {Asplund}}]{2001ApJ...556L..63A}
{Allende Prieto}, C., {Lambert}, D.~L., \& {Asplund}, M. 2001, \apjl, 556, L63

\bibitem[{{Biazzo} {et~al.}(2015){Biazzo}, {Gratton}, {Desidera}, {Lucatello},
  {Sozzetti}, {Bonomo}, {Damasso}, {Gandolfi}, {Affer}, {Boccato}, {Borsa},
  {Claudi}, {Cosentino}, {Covino}, {Knapic}, {Lanza}, {Maldonado}, {Marzari},
  {Micela}, {Molaro}, {Pagano}, {Pedani}, {Pillitteri}, {Piotto}, {Poretti},
  {Rainer}, {Santos}, {Scandariato}, \& {Zanmar Sanchez}}]{2015A&A...583A.135B}
{Biazzo}, K., {Gratton}, R., {Desidera}, S., {et~al.} 2015, \aap, 583, A135

\bibitem[{{Bonfanti} {et~al.}(2015){Bonfanti}, {Ortolani}, {Piotto}, \&
  {Nascimbeni}}]{2015A&A...575A..18B}
{Bonfanti}, A., {Ortolani}, S., {Piotto}, G., \& {Nascimbeni}, V. 2015, \aap,
  575, A18

\bibitem[{{Bressan} {et~al.}(2012){Bressan}, {Marigo}, {Girardi}, {Salasnich},
  {Dal Cero}, {Rubele}, \& {Nanni}}]{2012MNRAS.427..127B}
{Bressan}, A., {Marigo}, P., {Girardi}, L., {et~al.} 2012, \mnras, 427, 127

\bibitem[{{Buchhave} \& {Latham}(2015)}]{2015ApJ...808..187B}
{Buchhave}, L.~A. \& {Latham}, D.~W. 2015, \apj, 808, 187

\bibitem[{{Buchhave} {et~al.}(2012){Buchhave}, {Latham}, {Johansen},
  {Bizzarro}, {Torres}, {Rowe}, {Batalha}, {Borucki}, {Brugamyer}, {Caldwell},
  {Bryson}, {Ciardi}, {Cochran}, {Endl}, {Esquerdo}, {Ford}, {Geary},
  {Gilliland}, {Hansen}, {Isaacson}, {Laird}, {Lucas}, {Marcy}, {Morse},
  {Robertson}, {Shporer}, {Stefanik}, {Still}, \&
  {Quinn}}]{2012Natur.486..375B}
{Buchhave}, L.~A., {Latham}, D.~W., {Johansen}, A., {et~al.} 2012, \nat, 486,
  375

\bibitem[{{Casagrande} {et~al.}(2011){Casagrande}, {Sch{\"o}nrich}, {Asplund},
  {Cassisi}, {Ram{\'{\i}}rez}, {Mel{\'e}ndez}, {Bensby}, \&
  {Feltzing}}]{2011A&A...530A.138C}
{Casagrande}, L., {Sch{\"o}nrich}, R., {Asplund}, M., {et~al.} 2011, \aap, 530,
  A138

\bibitem[{{Cassan} {et~al.}(2012){Cassan}, {Kubas}, {Beaulieu}, {Dominik},
  {Horne}, {Greenhill}, {Wambsganss}, {Menzies}, {Williams}, {J{\o}rgensen},
  {Udalski}, {Bennett}, {Albrow}, {Batista}, {Brillant}, {Caldwell}, {Cole},
  {Coutures}, {Cook}, {Dieters}, {Prester}, {Donatowicz}, {Fouqu{\'e}}, {Hill},
  {Kains}, {Kane}, {Marquette}, {Martin}, {Pollard}, {Sahu}, {Vinter},
  {Warren}, {Watson}, {Zub}, {Sumi}, {Szyma{\'n}ski}, {Kubiak}, {Poleski},
  {Soszynski}, {Ulaczyk}, {Pietrzy{\'n}ski}, \&
  {Wyrzykowski}}]{2012Natur.481..167C}
{Cassan}, A., {Kubas}, D., {Beaulieu}, J.-P., {et~al.} 2012, \nat, 481, 167

\bibitem[{{Curran}(2014)}]{2014arXiv1411.3816C}
{Curran}, P.~A. 2014, ArXiv e-prints [\eprint[arXiv]{1411.3816}]

\bibitem[{{da Silva} {et~al.}(2006){da Silva}, {Girardi}, {Pasquini},
  {Setiawan}, {von der L{\"u}he}, {de Medeiros}, {Hatzes}, {D{\"o}llinger}, \&
  {Weiss}}]{2006A&A...458..609D}
{da Silva}, L., {Girardi}, L., {Pasquini}, L., {et~al.} 2006, \aap, 458, 609

\bibitem[{{da Silva} {et~al.}(2011){da Silva}, {Milone}, \&
  {Reddy}}]{2011A&A...526A..71D}
{da Silva}, R., {Milone}, A.~C., \& {Reddy}, B.~E. 2011, \aap, 526, A71

\bibitem[{{da Silva} {et~al.}(2015){da Silva}, {Milone}, \&
  {Rocha-Pinto}}]{2015A&A...580A..24D}
{da Silva}, R., {Milone}, A.~d.~C., \& {Rocha-Pinto}, H.~J. 2015, \aap, 580,
  A24

\bibitem[{{Delgado Mena} {et~al.}(2010){Delgado Mena}, {Israelian},
  {Gonz{\'a}lez Hern{\'a}ndez}, {Bond}, {Santos}, {Udry}, \&
  {Mayor}}]{2010ApJ...725.2349D}
{Delgado Mena}, E., {Israelian}, G., {Gonz{\'a}lez Hern{\'a}ndez}, J.~I.,
  {et~al.} 2010, \apj, 725, 2349

\bibitem[{{D{\"o}llinger} {et~al.}(2009){D{\"o}llinger}, {Hatzes}, {Pasquini},
  {Guenther}, \& {Hartmann}}]{2009A&A...505.1311D}
{D{\"o}llinger}, M.~P., {Hatzes}, A.~P., {Pasquini}, L., {Guenther}, E.~W., \&
  {Hartmann}, M. 2009, \aap, 505, 1311

\bibitem[{{Fischer} \& {Valenti}(2005)}]{2005ApJ...622.1102F}
{Fischer}, D.~A. \& {Valenti}, J. 2005, \apj, 622, 1102

\bibitem[{{Frandsen} \& {Lindberg}(1999)}]{1999anot.conf...71F}
{Frandsen}, S. \& {Lindberg}, B. 1999, in Astrophysics with the NOT, ed.
  H.~{Karttunen} \& V.~{Piirola}, 71

\bibitem[{{Gaidos}(2015)}]{2015ApJ...804...40G}
{Gaidos}, E. 2015, \apj, 804, 40

\bibitem[{{Ghezzi} {et~al.}(2010{\natexlab{a}}){Ghezzi}, {Cunha}, {Schuler}, \&
  {Smith}}]{2010ApJ...725..721G}
{Ghezzi}, L., {Cunha}, K., {Schuler}, S.~C., \& {Smith}, V.~V.
  2010{\natexlab{a}}, \apj, 725, 721

\bibitem[{{Ghezzi} {et~al.}(2010{\natexlab{b}}){Ghezzi}, {Cunha}, {Smith}, {de
  Ara{\'u}jo}, {Schuler}, \& {de la Reza}}]{2010ApJ...720.1290G}
{Ghezzi}, L., {Cunha}, K., {Smith}, V.~V., {et~al.} 2010{\natexlab{b}}, \apj,
  720, 1290

\bibitem[{{Ghezzi} \& {Johnson}(2015)}]{2015ApJ...812...96G}
{Ghezzi}, L. \& {Johnson}, J.~A. 2015, \apj, 812, 96

\bibitem[{{Girardi} {et~al.}(2000){Girardi}, {Bressan}, {Bertelli}, \&
  {Chiosi}}]{2000A&AS..141..371G}
{Girardi}, L., {Bressan}, A., {Bertelli}, G., \& {Chiosi}, C. 2000, \aaps, 141,
  371

\bibitem[{{Gonzalez}(1997)}]{1997MNRAS.285..403G}
{Gonzalez}, G. 1997, \mnras, 285, 403

\bibitem[{{Gonzalez}(2011)}]{2011MNRAS.416L..80G}
{Gonzalez}, G. 2011, \mnras, 416, L80

\bibitem[{{Gonz{\'a}lez Hern{\'a}ndez} {et~al.}(2013){Gonz{\'a}lez
  Hern{\'a}ndez}, {Delgado-Mena}, {Sousa}, {Israelian}, {Santos}, {Adibekyan},
  \& {Udry}}]{2013A&A...552A...6G}
{Gonz{\'a}lez Hern{\'a}ndez}, J.~I., {Delgado-Mena}, E., {Sousa}, S.~G.,
  {et~al.} 2013, \aap, 552, A6

\bibitem[{{Gonz{\'a}lez Hern{\'a}ndez} {et~al.}(2010){Gonz{\'a}lez
  Hern{\'a}ndez}, {Israelian}, {Santos}, {Sousa}, {Delgado-Mena}, {Neves}, \&
  {Udry}}]{2010ApJ...720.1592G}
{Gonz{\'a}lez Hern{\'a}ndez}, J.~I., {Israelian}, G., {Santos}, N.~C., {et~al.}
  2010, \apj, 720, 1592

\bibitem[{{Gratton} {et~al.}(2001){Gratton}, {Bonanno}, {Bruno}, {Cali},
  {Claudi}, {Cosentino}, {Desidera}, {Diego}, {Farisato}, {Martorana},
  {Rebeschini}, \& {Scuderi}}]{sarg}
{Gratton}, R.~G., {Bonanno}, G., {Bruno}, P., {et~al.} 2001, Experimental
  Astronomy, 12, 107

\bibitem[{{Haywood}(2009)}]{2009ApJ...698L...1H}
{Haywood}, M. 2009, \apjl, 698, L1

\bibitem[{{Hekker} \& {Mel{\'e}ndez}(2007)}]{2007A&A...475.1003H}
{Hekker}, S. \& {Mel{\'e}ndez}, J. 2007, \aap, 475, 1003

\bibitem[{{Howard} {et~al.}(2013){Howard}, {Sanchis-Ojeda}, {Marcy}, {Johnson},
  {Winn}, {Isaacson}, {Fischer}, {Fulton}, {Sinukoff}, \&
  {Fortney}}]{2013Natur.503..381H}
{Howard}, A.~W., {Sanchis-Ojeda}, R., {Marcy}, G.~W., {et~al.} 2013, \nat, 503,
  381

\bibitem[{{Jofr{\'e}} {et~al.}(2015){Jofr{\'e}}, {Petrucci}, {Saffe}, {Saker},
  {de la Villarmois}, {Chavero}, {G{\'o}mez}, \& {Mauas}}]{2015A&A...574A..50J}
{Jofr{\'e}}, E., {Petrucci}, R., {Saffe}, C., {et~al.} 2015, \aap, 574, A50

\bibitem[{{Johnson} {et~al.}(2011){Johnson}, {Clanton}, {Howard}, {Bowler},
  {Henry}, {Marcy}, {Crepp}, {Endl}, {Cochran}, {MacQueen}, {Wright}, \&
  {Isaacson}}]{2011ApJS..197...26J}
{Johnson}, J.~A., {Clanton}, C., {Howard}, A.~W., {et~al.} 2011, \apjs, 197, 26

\bibitem[{{Kennedy} \& {Kenyon}(2008)}]{2008ApJ...673..502K}
{Kennedy}, G.~M. \& {Kenyon}, S.~J. 2008, \apj, 673, 502

\bibitem[{{Kjeldsen} \& {Bedding}(1995)}]{1995A&A...293...87K}
{Kjeldsen}, H. \& {Bedding}, T.~R. 1995, \aap, 293 [\eprint{astro-ph/9403015}]

\bibitem[{{Kurucz}(1993)}]{1993KurCD..13.....K}
{Kurucz}, R. 1993, ATLAS9 Stellar Atmosphere Programs and 2 km/s grid.~Kurucz
  CD-ROM No.~13.~ Cambridge, Mass.: Smithsonian Astrophysical Observatory,
  1993., 13

\bibitem[{{Lemasle} {et~al.}(2008){Lemasle}, {Fran{\c c}ois}, {Piersimoni},
  {Pedicelli}, {Bono}, {Laney}, {Primas}, \&
  {Romaniello}}]{2008A&A...490..613L}
{Lemasle}, B., {Fran{\c c}ois}, P., {Piersimoni}, A., {et~al.} 2008, \aap, 490,
  613

\bibitem[{{Lloyd}(2011)}]{2011ApJ...739L..49L}
{Lloyd}, J.~P. 2011, \apjl, 739, L49

\bibitem[{{Lloyd}(2013)}]{2013ApJ...774L...2L}
{Lloyd}, J.~P. 2013, \apjl, 774, L2

\bibitem[{{Lodders}(2003)}]{2003ApJ...591.1220L}
{Lodders}, K. 2003, \apj, 591, 1220

\bibitem[{{Maldonado} {et~al.}(2015){Maldonado}, {Eiroa}, {Villaver},
  {Montesinos}, \& {Mora}}]{2015A&A...579A..20M}
{Maldonado}, J., {Eiroa}, C., {Villaver}, E., {Montesinos}, B., \& {Mora}, A.
  2015, \aap, 579, A20

\bibitem[{{Maldonado} {et~al.}(2013){Maldonado}, {Villaver}, \&
  {Eiroa}}]{2013A&A...554A..84M}
{Maldonado}, J., {Villaver}, E., \& {Eiroa}, C. 2013, \aap, 554, A84

\bibitem[{{Massarotti} {et~al.}(2008){Massarotti}, {Latham}, {Stefanik}, \&
  {Fogel}}]{2008AJ....135..209M}
{Massarotti}, A., {Latham}, D.~W., {Stefanik}, R.~P., \& {Fogel}, J. 2008, \aj,
  135, 209

\bibitem[{{Mayor} {et~al.}(2011){Mayor}, {Marmier}, {Lovis}, {Udry},
  {S{\'e}gransan}, {Pepe}, {Benz}, {Bertaux}, {Bouchy}, {Dumusque}, {Lo Curto},
  {Mordasini}, {Queloz}, \& {Santos}}]{2011arXiv1109.2497M}
{Mayor}, M., {Marmier}, M., {Lovis}, C., {et~al.} 2011, ArXiv e-prints
  [\eprint[arXiv]{1109.2497}]

\bibitem[{{Mel{\'e}ndez} {et~al.}(2009){Mel{\'e}ndez}, {Asplund}, {Gustafsson},
  \& {Yong}}]{2009ApJ...704L..66M}
{Mel{\'e}ndez}, J., {Asplund}, M., {Gustafsson}, B., \& {Yong}, D. 2009, \apjl,
  704, L66

\bibitem[{{Mortier} {et~al.}(2013){Mortier}, {Santos}, {Sousa}, {Adibekyan},
  {Delgado Mena}, {Tsantaki}, {Israelian}, \& {Mayor}}]{2013A&A...557A..70M}
{Mortier}, A., {Santos}, N.~C., {Sousa}, S.~G., {et~al.} 2013, \aap, 557, A70

\bibitem[{{Mustill} \& {Villaver}(2012)}]{2012ApJ...761..121M}
{Mustill}, A.~J. \& {Villaver}, E. 2012, \apj, 761, 121

\bibitem[{{Niedzielski} {et~al.}(2016){Niedzielski}, {Deka-Szymankiewicz},
  {Adamczyk}, {Adam{\'o}w}, {Nowak}, \& {Wolszczan}}]{2016A&A...585A..73N}
{Niedzielski}, A., {Deka-Szymankiewicz}, B., {Adamczyk}, M., {et~al.} 2016,
  \aap, 585, A73

\bibitem[{{Nissen}(2015)}]{2015A&A...579A..52N}
{Nissen}, P.~E. 2015, \aap, 579, A52

\bibitem[{{Pasquini} {et~al.}(2007){Pasquini}, {D{\"o}llinger}, {Weiss},
  {Girardi}, {Chavero}, {Hatzes}, {da Silva}, \&
  {Setiawan}}]{2007A&A...473..979P}
{Pasquini}, L., {D{\"o}llinger}, M.~P., {Weiss}, A., {et~al.} 2007, \aap, 473,
  979

\bibitem[{{Pfeiffer} {et~al.}(1998){Pfeiffer}, {Frank}, {Baumueller},
  {Fuhrmann}, \& {Gehren}}]{foces}
{Pfeiffer}, M.~J., {Frank}, C., {Baumueller}, D., {Fuhrmann}, K., \& {Gehren},
  T. 1998, \aaps, 130, 381

\bibitem[{{Ram{\'{\i}}rez} {et~al.}(2010){Ram{\'{\i}}rez}, {Asplund},
  {Baumann}, {Mel{\'e}ndez}, \& {Bensby}}]{2010A&A...521A..33R}
{Ram{\'{\i}}rez}, I., {Asplund}, M., {Baumann}, P., {Mel{\'e}ndez}, J., \&
  {Bensby}, T. 2010, \aap, 521, A33

\bibitem[{{Ram{\'{\i}}rez} {et~al.}(2015){Ram{\'{\i}}rez}, {Khanal}, {Aleo},
  {Sobotka}, {Liu}, {Casagrande}, {Mel{\'e}ndez}, {Yong}, {Lambert}, \&
  {Asplund}}]{2015ApJ...808...13R}
{Ram{\'{\i}}rez}, I., {Khanal}, S., {Aleo}, P., {et~al.} 2015, \apj, 808, 13

\bibitem[{{Ram{\'{\i}}rez} {et~al.}(2009){Ram{\'{\i}}rez}, {Mel{\'e}ndez}, \&
  {Asplund}}]{2009A&A...508L..17R}
{Ram{\'{\i}}rez}, I., {Mel{\'e}ndez}, J., \& {Asplund}, M. 2009, \aap, 508, L17

\bibitem[{{Ram{\'{\i}}rez} {et~al.}(2014){Ram{\'{\i}}rez}, {Mel{\'e}ndez}, \&
  {Asplund}}]{2014A&A...561A...7R}
{Ram{\'{\i}}rez}, I., {Mel{\'e}ndez}, J., \& {Asplund}, M. 2014, \aap, 561, A7

\bibitem[{{Raskin} {et~al.}(2011){Raskin}, {van Winckel}, {Hensberge},
  {Jorissen}, {Lehmann}, {Waelkens}, {Avila}, {de Cuyper}, {Degroote},
  {Dubosson}, {Dumortier}, {Fr{\'e}mat}, {Laux}, {Michaud}, {Morren}, {Perez
  Padilla}, {Pessemier}, {Prins}, {Smolders}, {van Eck}, \&
  {Winkler}}]{2011A&A...526A..69R}
{Raskin}, G., {van Winckel}, H., {Hensberge}, H., {et~al.} 2011, \aap, 526, A69

\bibitem[{{Reffert} {et~al.}(2015){Reffert}, {Bergmann}, {Quirrenbach},
  {Trifonov}, \& {K{\"u}nstler}}]{2015A&A...574A.116R}
{Reffert}, S., {Bergmann}, C., {Quirrenbach}, A., {Trifonov}, T., \&
  {K{\"u}nstler}, A. 2015, \aap, 574, A116

\bibitem[{{Sadakane} {et~al.}(2005){Sadakane}, {Ohnishi}, {Ohkubo}, \&
  {Takeda}}]{2005PASJ...57..127S}
{Sadakane}, K., {Ohnishi}, T., {Ohkubo}, M., \& {Takeda}, Y. 2005, \pasj, 57,
  127

\bibitem[{{Santos} {et~al.}(2004){Santos}, {Israelian}, \&
  {Mayor}}]{2004A&A...415.1153S}
{Santos}, N.~C., {Israelian}, G., \& {Mayor}, M. 2004, \aap, 415, 1153

\bibitem[{{Schlaufman} \& {Winn}(2013)}]{2013ApJ...772..143S}
{Schlaufman}, K.~C. \& {Winn}, J.~N. 2013, \apj, 772, 143

\bibitem[{{Schuler} {et~al.}(2005){Schuler}, {Kim}, {Tinker}, {King}, {Hatzes},
  \& {Guenther}}]{2005ApJ...632L.131S}
{Schuler}, S.~C., {Kim}, J.~H., {Tinker}, Jr., M.~C., {et~al.} 2005, \apjl,
  632, L131

\bibitem[{{Sneden}(1973)}]{1973PhDT.......180S}
{Sneden}, C.~A. 1973, PhD thesis, THE UNIVERSITY OF TEXAS AT AUSTIN.

\bibitem[{{Sousa} {et~al.}(2011){Sousa}, {Santos}, {Israelian}, {Mayor}, \&
  {Udry}}]{2011A&A...533A.141S}
{Sousa}, S.~G., {Santos}, N.~C., {Israelian}, G., {Mayor}, M., \& {Udry}, S.
  2011, \aap, 533, A141

\bibitem[{{Spina} {et~al.}(2016){Spina}, {Mel{\'e}ndez}, \&
  {Ram{\'{\i}}rez}}]{2015arXiv151101012S}
{Spina}, L., {Mel{\'e}ndez}, J., \& {Ram{\'{\i}}rez}, I. 2016, \aap, 585, A152

\bibitem[{{Takeda} {et~al.}(2005){Takeda}, {Ohkubo}, {Sato}, {Kambe}, \&
  {Sadakane}}]{2005PASJ...57...27T}
{Takeda}, Y., {Ohkubo}, M., {Sato}, B., {Kambe}, E., \& {Sadakane}, K. 2005,
  \pasj, 57, 27

\bibitem[{{Takeda} {et~al.}(2008){Takeda}, {Sato}, \&
  {Murata}}]{2008PASJ...60..781T}
{Takeda}, Y., {Sato}, B., \& {Murata}, D. 2008, \pasj, 60, 781

\bibitem[{{Thiabaud} {et~al.}(2015){Thiabaud}, {Marboeuf}, {Alibert}, {Leya},
  \& {Mezger}}]{2015A&A...580A..30T}
{Thiabaud}, A., {Marboeuf}, U., {Alibert}, Y., {Leya}, I., \& {Mezger}, K.
  2015, \aap, 580, A30

\bibitem[{{Villaver} \& {Livio}(2009)}]{2009ApJ...705L..81V}
{Villaver}, E. \& {Livio}, M. 2009, \apjl, 705, L81

\bibitem[{{Villaver} {et~al.}(2014){Villaver}, {Livio}, {Mustill}, \&
  {Siess}}]{2014ApJ...794....3V}
{Villaver}, E., {Livio}, M., {Mustill}, A.~J., \& {Siess}, L. 2014, \apj, 794,
  3

\bibitem[{{Wang} \& {Fischer}(2015)}]{2015AJ....149...14W}
{Wang}, J. \& {Fischer}, D.~A. 2015, \aj, 149, 14

\bibitem[{{Zieli{\'n}ski} {et~al.}(2012){Zieli{\'n}ski}, {Niedzielski},
  {Wolszczan}, {Adam{\'o}w}, \& {Nowak}}]{2012A&A...547A..91Z}
{Zieli{\'n}ski}, P., {Niedzielski}, A., {Wolszczan}, A., {Adam{\'o}w}, M., \&
  {Nowak}, G. 2012, \aap, 547, A91

\end{thebibliography}

\begin{appendix}
\section{Abundance ratios as a function of stellar metallicity}
\label{apendiceA}
\begin{figure*}
\centering
\includegraphics[angle=270,scale=0.45]{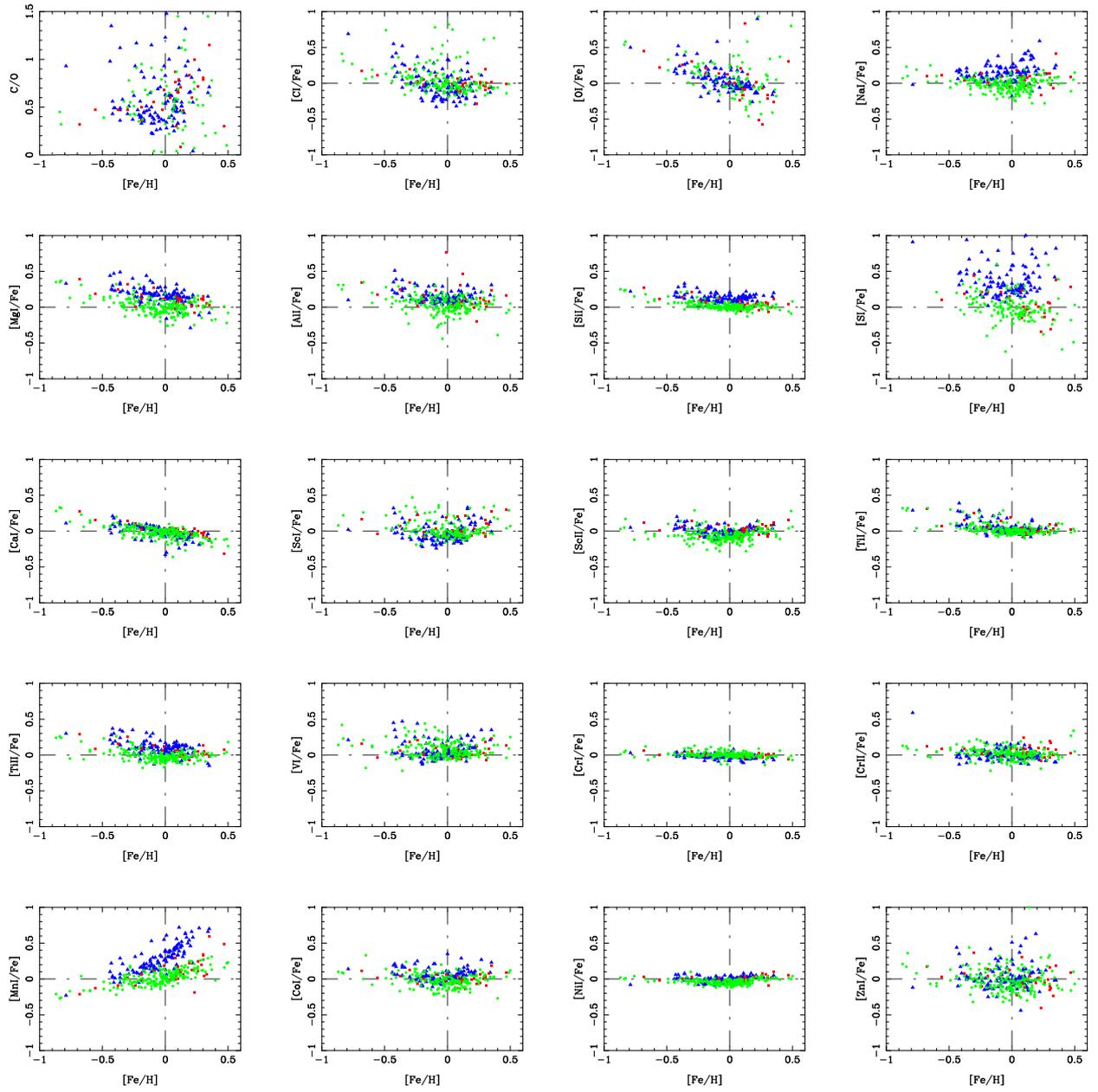}
\caption{C/O and chemical abundance ratios of [X/Fe] as a function of the stellar metallicity.
 MS stars are plotted as green circles, subgiants as red squares, and giants as blue triangles. 
}
\label{swds_swods_xfe_vs_feh}
\end{figure*}

\end{appendix}

\Online
\section*{Online material}

  Results produced in the framework of this work
  are only available
  in the electronic version of the corresponding paper or
  at the CDS via anonymous ftp to cdsarc.u-strasbg.fr (130.79.128.5)
  or via {\tt http://cdsweb.u-strasbg.fr/cgi-bin/qcat?J/A+A/}

  Table~\ref{parameters_table_full} lists all the stars analysed in this work.
  Note that the data for the majority of the main-sequence stars comes
  from MA15 and is not reproduced here.
  The table provides:
  HIP number (column 1); HD number (column 2);
  effective temperature in kelvin (column 3); logarithm of the surface gravity in cms$^{\rm -2}$ (column 4);
  microturbulent velocity in kms$^{\rm -1}$ (column 5); final metallicity in dex (column 6);
  spectrograph (column 7); stellar age in Gyr (column 8); stellar mass in solar units (column 9);
  and stellar radii in solar units (column 10).
  Each measured quantity is accompanied by its corresponding uncertainty.

  Table~\ref{abundance_table_full}   gives the
  abundances of C~{\sc I}, O~{\sc I}, Na~{\sc I}, Mg~{\sc I}, Al~{\sc I}, Si~{\sc I},
  S~{\sc I}, Ca~{\sc I}, Sc~{\sc I}, Sc~{\sc II}, Ti~{\sc I}, Ti~{\sc II},
  V~{\sc I} (HFS taken into account), Cr~{\sc I},  Cr~{\sc II}
  Mn~{\sc I}, Co~{\sc I} (HFS taken into account) , Ni~{\sc I}, 
  and Zn~{\sc I} 
  They are expressed relative to the
  solar value, i.e
  $[X/H]=\log(N_{X}/N_{H}) - \log(N_{X}/N_{H})_{\odot}$.
  For each star abundances are given in the first row, whilst uncertainties
  are given in the second row.
  Note that the data for the majority of the main-sequence stars comes
  from MA15 and is not reproduced here.
  Abundances of carbon and oxygen of the stars from
  M15 were recomputed as described in Section~\ref{secction_observations}.
  They are given in Table~\ref{carbon_oxygen_MA15_stars}.


\end{document}